\documentclass[aps,prl,reprint,secnumarabic,superscriptaddress]{revtex4-2}

\usepackage{graphicx}
\usepackage{amsmath,amssymb,bm}
\usepackage{braket}
\usepackage{dcolumn}
\usepackage{hyperref}
\usepackage{array}
\usepackage{multirow}
\usepackage{longtable}
\usepackage{booktabs}
\usepackage{xcolor}
\usepackage{float}
 \usepackage{bm}
\usepackage{siunitx}
\begin{document}

\title{Towards the First Realization of Weyl Magnons in the Non-Coplanar Antiferromagnet MnTe$_2$}


\author{Ahmed E. Fahmy}
\thanks{These authors contributed equally to this work.}
\affiliation{Department of Physics, The Ohio State University, Columbus, OH 43210, USA}

\author{Archibald J. Williams}
\thanks{These authors contributed equally to this work.}
\affiliation{Department of Chemistry and Biochemistry, The Ohio State University, Columbus, OH 43210, USA}

\author{Yufei Li}
\thanks{These authors contributed equally to this work.}
\affiliation{Department of Physics, The Ohio State University, Columbus, OH 43210, USA}

\author{Thuc T. Mai}
\affiliation{Quantum Measurements Division, Physical Measurement Laboratory, National Institute of Standards and Technology, Gaithersburg, MD 20899, USA}

\author{Kevin F. Garrity}
\affiliation{Materials Measurement Science Division, Material Measurement Laboratory, National Institute of Standards and Technology, Gaithersburg, MD 20899, USA}

\author{Matthew B. Stone}
\affiliation{Neutron Scattering Division, Oak Ridge National Laboratory, Oak Ridge, Tennessee 37831, USA}

\author{Mohammed J. Karaki}
\affiliation{Department of Physics, The Ohio State University, Columbus, OH 43210, USA}

\author{Sara Haravifard}
\affiliation{Department of Physics, Duke University, Durham, North Carolina 27708, USA}
\affiliation{Department of Mechanical Engineering and Materials Science, Duke University, Durham, North Carolina 27708, USA}

\author{Angela R. Hight Walker}
\affiliation{Quantum Measurements Division, Physical Measurement Laboratory, National Institute of Standards and Technology, Gaithersburg, MD 20899, USA}

\author{Rolando Valdés Aguilar}
\email{rvaldesag@gmail.com}
\affiliation{Department of Physics, The Ohio State University, Columbus, OH 43210, USA}

\author{Joshua E. Goldberger}
\email{goldberger.4@osu.edu}
\affiliation{Department of Chemistry and Biochemistry, The Ohio State University, Columbus, OH 43210, USA}

\author{Yuan-Ming Lu}
\email{lu.1435@osu.edu}
\affiliation{Department of Physics, The Ohio State University, Columbus, OH 43210, USA}


\begin{abstract}
Using a combination of band representation analysis, inelastic neutron scattering (INS), magneto-Raman spectroscopy measurements, and linear spin wave theory, we establish that the non-coplanar antiferromagnet MnTe$_2$ is a tunable Weyl magnon material, hosting symmetry-protected topological nodal lines in its magnon band structure, protected by the the non-coplanar nature of the antiferromagnetic ordering, that transition into Weyl magnons upon the application of symmetry-breaking perturbations using an external magnetic field. By constructing a spin model that reproduces the observed INS magnon spectra and field-dependence of the Raman $\Gamma$-magnons, we directly probe the topological magnon nodal lines and observe their associated signature of non-trivial topology through the pseudo-spin winding of the scattering intensity in angular scans near the nodal lines. Finally, we discuss the emergence of Weyl magnons in the spectrum through an external magnetic field, and analyze the associated magnon thermal Hall effect allowed by the cubic-symmetry breaking and the associated magnon surface modes, shedding light on future in-field INS and thermal Hall experiments. This work establishes a clear magnonic analog to Weyl electrons, enabling further exploration of topological behavior in bosonic systems and highlighting the interplay between magnetic order and band topology in non-coplanar antiferromagnets.
\end{abstract}

\maketitle


In magnetically ordered materials, low-energy spin excitations manifest as quantized collective modes known as magnons. These quasiparticles are the chargeless bosonic analogues of electrons in solids, can propagate through the lattice, and have well-defined energy bands in momentum space. When these magnon bands acquire non-trivial topological characteristics—such as Berry curvature, Chern number, or protected edge states—the resulting excitations are termed topological magnons \cite{Shindou2013, Mook2014, McClarty2021}. These excitations are protected by the underlying symmetries and topology of the magnetic structure, making them robust against elastic backscattering from defects or disorder \cite{MurakamiPhysRevB.87.174427,HalperinPhysRevB.25.2185}. Owing to their bosonic statistics and intrinsic magnetic moment, magnons can transport angular momentum without electric charge, offering a potential platform for low-dissipation spin transport \cite{MagnonTransportInYIGCornelissen_2015,Chumak2015}. The confluence of topological protection and spin-carrying capacity has spurred proposals for a variety of topological magnonic devices, including chiral spin-wave waveguides, thermal Hall-based logic elements, and unidirectional magnon diodes \cite{Chumak2015, Nakata2017, Wang2018}. In particular, Weyl magnons\cite{AMookPhysRevLett.117.157204,Li2021weylmagnon,doi:10.1126/sciadv.ade7731,ExtraWeylMagnonPhysRevB.95.224403,ExtraWeylMagnon2Owerre_2018} are a striking subclass of topological magnons in which isolated, linear band crossings (Weyl points) appear in 3D momentum space, acting as monopoles of Berry curvature and enforcing nonzero Chern numbers on intermediate 2D slices of the Brillouin zone (BZ). Therefore, their surface projections are joined by open magnon‑arc modes \cite{doi:10.1126/sciadv.ade7731,karaki2024highthroughputsearchtopologicalmagnon}, and the resulting Berry‑curvature flux yields pronounced anomalous thermal Hall and chiral magnon transport features. A hallmark of these excitations is their tunability: Weyl
nodes can be created, shifted, or annihilated by external perturbations such as magnetic fields or
strain, offering an unusual degree of control over topological bosonic quasiparticles.


Quasiparticle Weyl excitations have been experimentally confirmed not only in electronic systems \cite{WeylFermionsExperimental1Xu_2015,WeylFermionsExperimental2PhysRevX.5.031013}, but also in artificial photonic crystals \cite{WeylPhotonsLu_2015} and phononic crystals \cite{WeylPhononsLi2018-ku}. However, no clear evidence for Weyl magnons has been found despite several theoretically-proposed spin models that predict Weyl magnons \cite{AMookPhysRevLett.117.157204,Li2021weylmagnon,doi:10.1126/sciadv.ade7731,ExtraWeylMagnonPhysRevB.95.224403,ExtraWeylMagnon2Owerre_2018}. From a symmetry perspective, Weyl electrons \cite{WeylFermionsTheory1PhysRevB.83.205101,WeylFermionsTheory2PhysRevLett.113.187202} are forbidden in the presence of a combined inversion $\mathcal{I}$ and time-reversal $\mathcal{T}$ symmetry that give rise to two-fold Kramers' degeneracy of bands. However, magnonic systems are fundamentally different; magnetic order breaks physical time reversal symmetry and magnons are integer spin, bosonic excitations that do not follow Kramers' theorem where $\mathcal{T}^2=+1$. This suggests that Weyl magnons will generally be allowed in magnonic band structures. However, even in the absence of time reversal symmetry, coplanar Heisenberg magnets have an effective time reversal symmetry $\bar{\mathcal{T}}$, a product of time reversal and a global $\pi-$spin rotation, that can prohibit Weyl magnons when the system is inversion-symmetric \cite{MookEffectiveTimeReversalPhysRevB.99.014427}. This results from the constraints that each symmetry places on the Berry curvature $\Omega_{n}(\vec{k})$ of magnon bands: inversion imposes $\Omega_{n}(\vec{k}) = \Omega_{n}(-\vec{k})$ and effective time reversal imposes  $\Omega_{n}(\vec{k}) = -\Omega_{n}(-\vec{k})$. Combined, they result in a vanishing Berry curvature. Since Weyl points are effectively monopole sources or sinks of Berry curvature, Weyl points are prohibited in the presence of both $\mathcal{I}$ and $\bar{\mathcal{T}}$ symmetries. This serves as the key motivation for choosing a non-coplanar, magnetically-ordered material without an effective time-reversal symmetry in order to realize Weyl magnons.

\begin{figure}[t]
    \centering
    \includegraphics[width=\columnwidth]{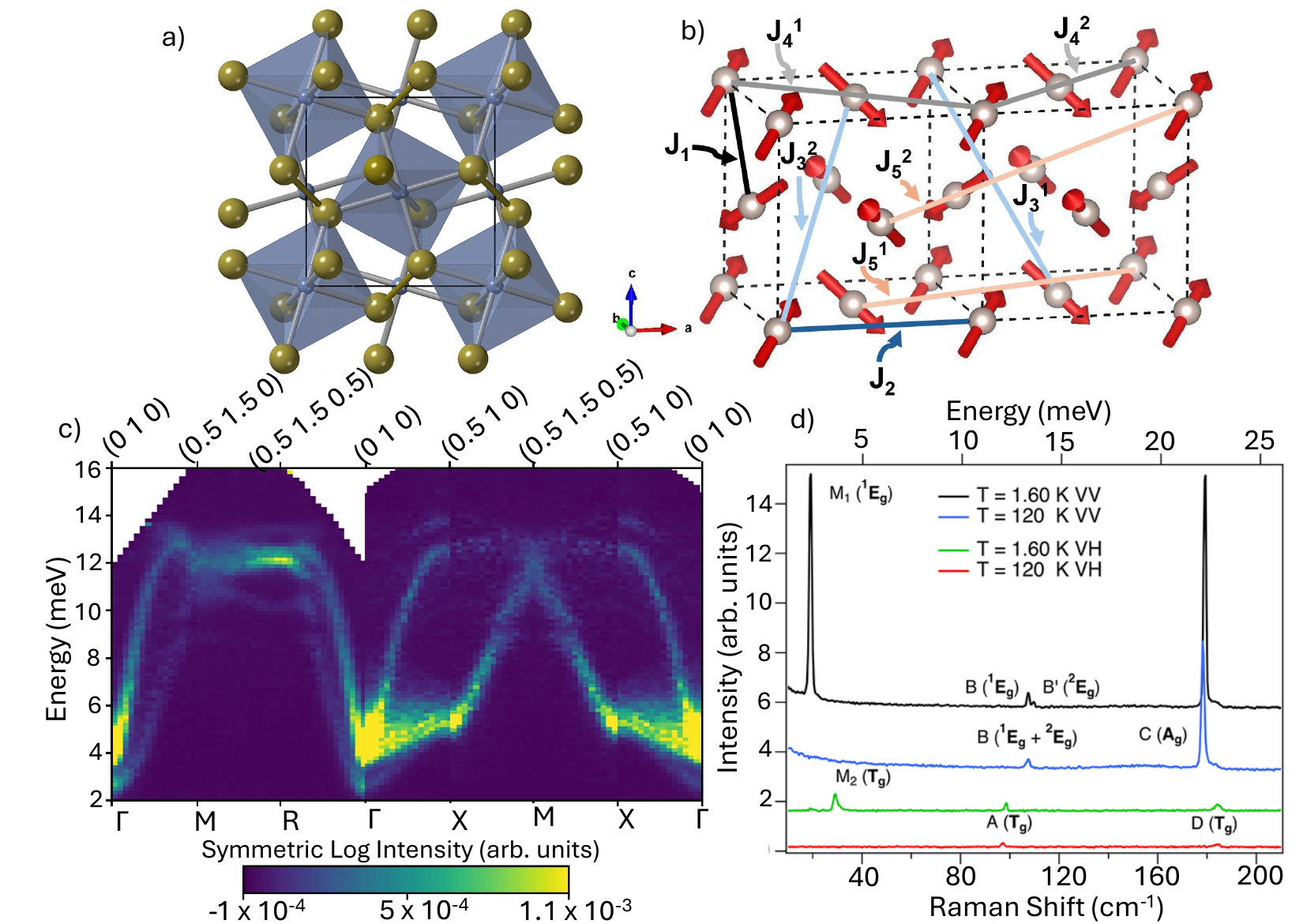}
    \caption{a) Crystal Structure of MnTe$_2$: blue and gold spheres represent the Mn and Te ions, respectively. Each Mn ion is bonded to six Te ions forming tilted corner-sharing octahedra, colored in blue gray. b) The magnetic structure of MnTe$_2$ highlighting the different $J$ couplings used for the LSWT fitting. The dotted lines indicates the magnetic unit cell and the coincident nuclear unit cell  c) Inelastic neutron scattering measurements of MnTe$_2$ at $5$ K along the principal directions in reciprocal space, with the different directions merged together. d) Temperature dependent Raman spectra above and below $T_N$, collected in both parallel (VV) and cross (VH) polarization configurations. The two magnons are labeled $M_1$ and $M_2$ while the four phonons are labeled A-D in increasing energy, with symmetries noted in parentheses.} 
    \label{Crystal and Magnetic Structure of MnTe2 Figure 1}
\end{figure}

While there have been previous reports on observations of topological magnons \cite{Chisnell2015,Chisnell2016,CoTiO3McClartyNatComm.10.1038/s41467-021-23851-0,CrI3PhysRevX.8.041028,ThermalEvolutionOfMagnonsinCrBr3PhysRevLett.129.127201,Yao_2018,Cu3TeO6DiracandTriplyDegBao_2018,GadoliniumMcClartyPhysRevLett.128.097201}, these have mainly been confined to hexagonal lattices, with collinear magnetic orders often involving Dirac crossings at high-symmetry points, or magnon Chern insulators. These examples provide limited generalization across broader material classes and geometries, restricting progress in scalable magnonic engineering. To realize the potential of topological magnonics in applications, it is crucial to first develop a unified theoretical and algorithmic framework capable of identifying diverse topological features across a wide range of magnetic materials. This has been accomplished recently \cite{karaki2024highthroughputsearchtopologicalmagnon} where it was shown how to use the symmetry of the magnetic structure and the atomic positions of the magnetic atoms to theoretically determine the topology of magnon bands. Here, we use those theoretical results and integrate them with experimental inelastic neutron scattering (INS), magneto-Raman scattering, and Hamiltonian modeling to prove the existence of topological magnon nodal lines. These gap out into tunable Weyl magnons when a magnetic field in a low-symmetry direction is applied. Direct evidence of these topological nodal lines in MnTe$_2$ are observed in zero-field INS through the angular dependence of the scattering intensity, revealing a characteristic pseudo-spin winding consistent with nontrivial magnon topology. These results establish MnTe$_2$ as a direct bosonic analogue of electronic Weyl systems, open new avenues for exploring topological excitations in non-coplanar magnetic insulators, and constitute a compelling experimental and theoretical demonstration of Weyl magnons. 


Among the recently developed catalog of topological magnon materials in Ref \cite{karaki2024highthroughputsearchtopologicalmagnon}, MnTe\textsubscript{2} was selected as a suitable target material to experimentally confirm the presence of topological Weyl magnons. It is an ideal candidate to confirm the presence of topological magnon bands due to its air stability, accessible transition temperature ($T_N \sim 87$ K), and electronic insulating behavior (energy gap $E_g = 0.87$ eV). MnTe\textsubscript{2} was also selected for its simple chemical formulation and the ease of large single-crystal growth (see Supplementary Sections \ref{crsytal-growth} \& \ref{magnetic-measurements}) appropriate for the measurement of topological magnons. MnTe\textsubscript{2}  crystallizes into the cubic pyrite structure in space group \textit{Pa$\overline{3}$} as illustrated in Fig. \ref{Crystal and Magnetic Structure of MnTe2 Figure 1} (a). The Mn atoms sit in corner sharing MnTe\textsubscript{6} octahedra and are further connected by Te\textsubscript{2}\textsuperscript{2-} dimers. This results in Mn\textsuperscript{2+} ions with S = $5/2$ on the atomic Wyckoff Position ($\mathcal{WP}$) $4a$ with site symmetry group of $\bar{3}$, where the $3$-fold axis points along the [$1 1 1$] crystal direction. Below the Néel temperature,  the Mn\textsuperscript{2+} moments order antiferromagnetically into a non-coplanar magnetic structure shown in Fig. \ref{Crystal and Magnetic Structure of MnTe2 Figure 1}(b) which can be described as four sublattices of Mn atoms with each Mn sublattice having moments that point along different local [$1 1 1$] directions, resulting in no net magnetic moment in the unit cell. 

\begin{figure*}[t]
 \hspace{-0.0cm}
 \centering
 \includegraphics[width=0.8\linewidth]{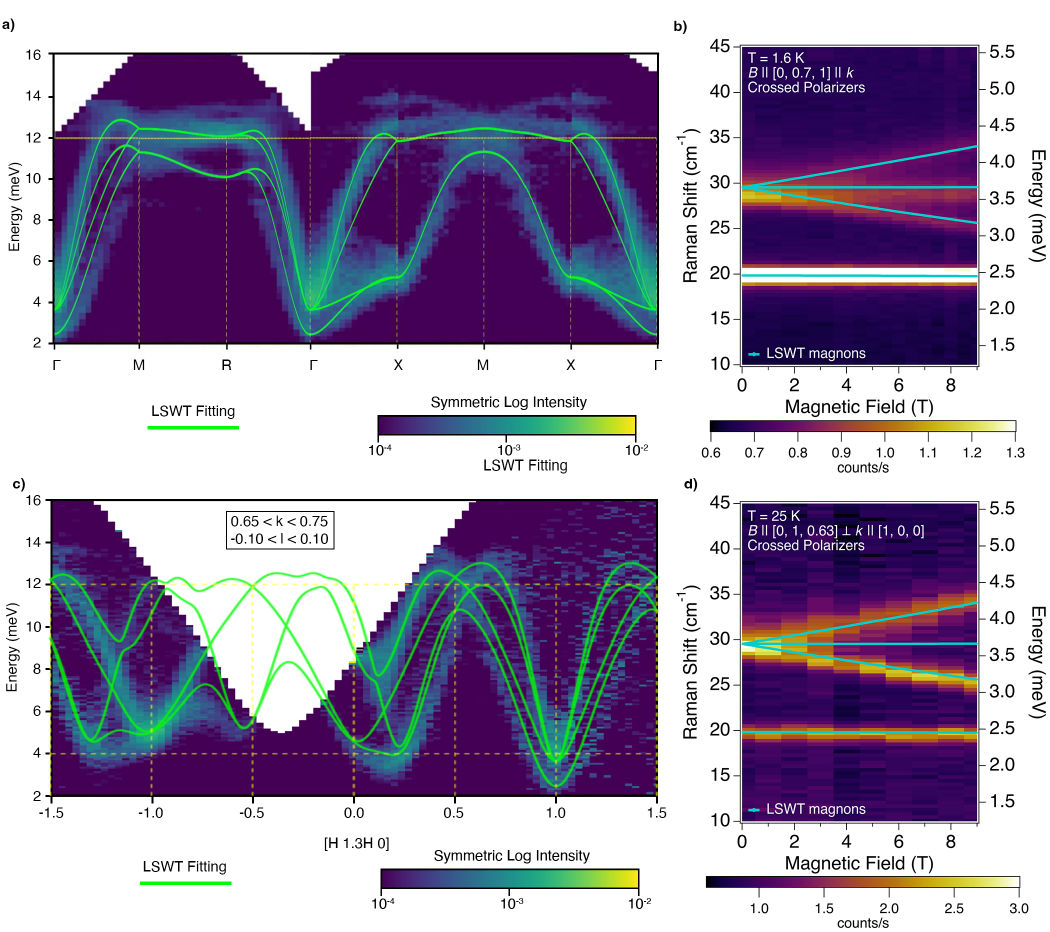}
 \caption{\textbf{Probing the magnon band structure of MnTe$_2$.} 
 a,c) An overlay of the LWST fitting of the magnonic spectra over the experimental INS data along (a) the high symmetry directions in the Brillouin zone and a (c) low symmetry direction $[K\; 0.2+1.3K\; 0]$. Refer to Supplementary Figs. \ref{comparisonmap1} \ref{comparisonmap2} \ref{comparisonmap3} \ref{comparisonmap4} for a more detailed comparison between LSWT simulations and INS measurements. b) Magneto-Raman data collected in the Faraday configuration with the field applied along [0 0.7 1] parallel to the light propagation direction. d) Magneto-Raman data collected in the Voigt configuration, with the field $B\parallel$ [0 1 0.63] and the light propagation along [1 0 0].}
 \label{figure2fittings}
 \end{figure*}

\begin{figure*}[t]
\centering
 \includegraphics[width=0.99\linewidth]{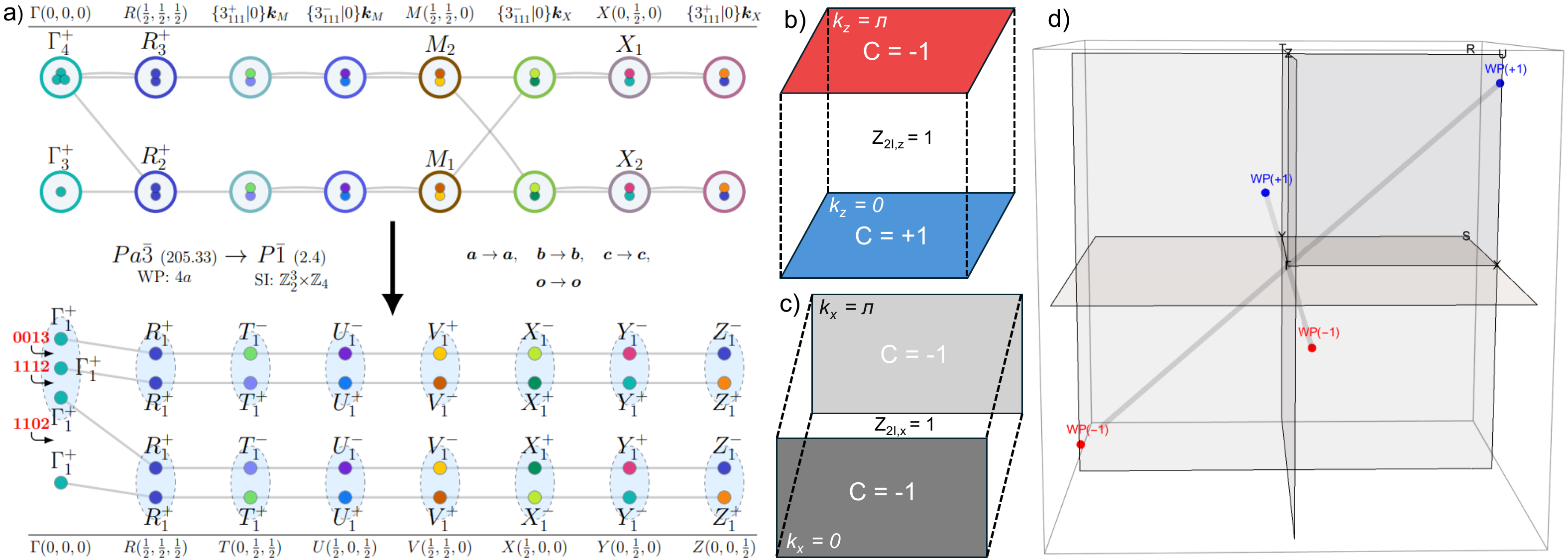}
 \caption{\textbf{Topological Weyl magnonic excitations in MnTe$_2$.} a) Schematic representation of the magnon band structure of MnTe$_2$ upon symmetry reduction from parent $\mathcal{MSG}$ $Pa \bar{3}$ into the subgroup $P\bar{1}$. The lower part manifests the explicit magnon band \textit{irreps} calculated using the obtained spin Hamiltonian. b-c) Demonstration of the calculated Chern numbers of the upper two bands over high-symmetry planes upon symmetry breaking, consistent with the $\mathcal{SI}$ predictions, thus requiring non-trivial bands to occur. d) Weyl points in k-space upon the application of a $14\;T$ magnetic field along the $[0\; 1\; 6]$ direction.}
 \label{figure3weylpointsinfield}
 \end{figure*}

Despite this magnetic ordering, the unit cell remains unchanged, resulting in a magnetic space group ($\mathcal{MSG}$)  of $Pa\bar{3}$. The magnon band representation induced from the $\mathcal{WP}$ 4a is:
\begin{equation*}
    \begin{aligned}
\left({ }^1 E_g\right)_{4 a} \uparrow P a \overline{3}(205.33)&= \; \Gamma_3^{+}(1) \oplus \Gamma_4^{+}(3) \oplus M_1(2) \oplus M_2(2) \\
& \; \; \oplus   R_2^{+}(2) \oplus R_3^{+}(2)  \oplus X_1(2) \oplus X_2(2)
\end{aligned}
\end{equation*}
where the $\mathcal{MSG}$ symmetries protect a three-fold degeneracy at the Brillouin zone center ($\Gamma$) of the $\Gamma_4^+$ magnon, and two doubly degenerate points at the $X,M,R$ momenta of the Brillouin zone, as verified by the combination of the INS and magneto-Raman measurements shown in Fig. \ref{figure2fittings}. This clearly demonstrates the compatibility relations of the magnon band \textit{irreps} (i.e. how magnon bands decompose and merge on the different high-symmetry manifolds of the BZ) schematically represented in the upper panel of Fig. \ref{figure3weylpointsinfield} (a). For example, on the line from $\Gamma$ to $R$, the triply-degenerate $\Gamma_4^+$ magnon at $\Gamma$ splits into three single bands and one of those combines with the singly-degenerate $\Gamma_3^+$ magnon to form a doubly degenerate magnon at $R$ where the other two bands also combine to form the other doubly-degenerate magnon, giving rise to the pair of two-fold degeneracies at $R$. These two doubly-degenerate magnon bands remain unbroken along the high-symmetry lines $R-M$ and $M-X$. Along $X-\Gamma$, each doubly-degenerate mode split into two single bands, and one of the single bands of one doubly-degenerate magnon combine with the two single bands of the other doubly-degenerate magnon to form the triply-degenerate $\Gamma$ magnon while the remaining band becomes the singly-degenerate $\Gamma$ mode. Additionally, an investigation of the compatibility relations on the $k_i=0$ high-symmetry planes reveals the existence of unavoidable nodal lines on these planes as demonstrated in the crossing between the $M(1/2,1/2,0)$ and $\{3^-_{111}|0 \}k_X$ momenta shown in Fig. \ref{figure3weylpointsinfield} (a) (see also Supplementary Section \ref{magnon-band-connectivity-supp-section}).

Weyl magnons, however, can be induced in the band structure of MnTe$_2$ upon the inclusion of certain symmetry-breaking perturbations as predicted through the symmetry-based algorithm \cite{karaki2024highthroughputsearchtopologicalmagnon} as following. Out of the $10$ subgroups of $Pa\Bar{3}$ that can be realized by applying external perturbations\cite{doi:10.1126/sciadv.ade7731}, only $5$ have nontrivial symmetry indicator ($\mathcal{SI}$) groups \cite{Peng2022}. Among those, only one subgroup, $P\Bar{1}$, host type-I topological magnons (i.e. guaranteed perturbation-induced magnon topology) \cite{karaki2024highthroughputsearchtopologicalmagnon}. This subgroup can be realized by a magnetic field along a low-symmetry direction which causes the little-group \textit{irreps} at the high-symmetry \textit{k}-points of $Pa\Bar{3}$ to decompose into \textit{irreps} of $P\Bar{1}$ (see Supplementary Section \ref{topo-of-middle-band-gap}).  The $\mathcal{SI}$ of $P\bar{1}$, with a $\mathbb{Z}_2^3 \times \mathbb{Z}_4$ group structure, can be calculated using the inversion symmetry eigenvalues at the $8$ high-symmetry momenta of $P\bar{1}$ \cite{Peng_2022}, and their possible values are shown in Supplementary Table \ref{symmetry-indicators-table} where they are written as a quadruplet set of the numbers $\mathbb{Z}_{2,x}\mathbb{Z}_{2,y}\mathbb{Z}_{2,z}\mathbb{Z}_4$. Upon this symmetry breaking, the different energetic ordering of magnon \textit{irreps} at the different high-symmetry momenta of $P\Bar{1}$ always ensures at least one magnon energy gap (specifically, the middle gap) with a topological $\mathcal{SI}$ \cite{karaki2024highthroughputsearchtopologicalmagnon}. The $\mathbb{Z}_{2,i=x,y,z}$ index diagnoses the Chern number \textit{modulo} $2$ on the $\mathbf{k}_i=0,\pi$ planes and the $\mathbb{Z}_4$ index diagnoses the difference between the Chern numbers on the two high-symmetry planes $\mathbf{k}_z=0,\pi$. Consequently, the $\mathcal{SI
 }$ value $1112$ of the second gap reveals an even number of perturbation-induced topologically nontrivial Weyl crossings in half of the Brillouin zone between the second and third magnon bands at generic momenta.

High-quality cm-sized single crystals of MnTe$_2$ were grown through a Bridgman process (Supplementary Fig. \ref{CrystalGrowth}) and confirmed to order antiferromagnetically below $87$ K (Supplementary Fig. \ref{magnetizationandsusceptibility}). INS experiments were performed to reveal the full magnon band structure. Fig. \ref{Crystal and Magnetic Structure of MnTe2 Figure 1} (c) illustrates the experimentally measureed magnon dispersion
along high symmetry directions as collected by single crystal INS. Measurements were collected at $5$ K with $22$ meV incident energy at the
SEQUOIA beamline of the Spallation Neutron Source. The zone center magnons energies are $2.5$ $\pm$ 0.2 meV  and $4.1$ $\pm$ 0.5 meV at zero magnetic field. Both
modes have a steep dispersion resulting in two broad doubly-degenerate modes, as indicated by the band representation analysis, from $\Gamma$ to $M$ that occur at $11.1$ $\pm$ 0.5 meV and $12.2$ $\pm$ 0.5 meV. At high momentum, from $M$ to $R$, two
low-intensity modes appear above and below a flat magnon dispersion along this direction. These modes are most likely phonon modes as the DFT calculations of the phonon dispersion show modes at these energies and momenta (details in the Supplementary Section \ref{DFT Phonons}) and as we show throughout the flow of the main text. In addition, these are seen only at high momentum transfer as expected given that, in a neutron experiment, phonons increase in intensity with momentum transfer $Q$ as $Q^2$ . Along the $\Gamma$ to $X$ direction, a crossover is seen as one band of the triply-degenerate state at $\Gamma$ joins
the lower energy mode while the other two modes remain degenerate and disperse to higher energies. At these high symmetry points of the
Brillouin zone, the degeneracies of the magnons follow the predictions of the symmetry analysis presented in the previous section (see Fig. \ref{figure3weylpointsinfield} (a)).

Raman experiments, performed at $1.6$ K and $120$ K, showed the presence of two magnon modes at the Brillouin zone center $\Gamma$ (Fig. \ref{Crystal and Magnetic Structure of MnTe2 Figure 1} (d), Supplementary Section \ref{Raman-tensors-polarization-temp}) in the magnetically ordered state and at zero magnetic field. At both temperatures four zone center phonons are identified at the energies of
$12.1\pm0.2$ meV, $13.3\pm0.2$ meV, $22.2\pm0.1$ meV, and $22.7\pm0.2$ meV, which match well with previous reports on MnTe$_2$ \cite{MULLER1991469}, and are labeled A through D in Fig. \ref{Crystal and Magnetic Structure of MnTe2 Figure 1}(d) with symmetries noted in parenthesis. Notably, the B mode, with \textit{irrep} $^1$E$_{\text{g}}$+$^2$E$_{\text{g}}$, seems to be split in the low temperature phase, where the two modes are labeled as B and B$'$. This splitting occurs because of the loss of time-reversal symmetry with the magnetic order removing the degeneracy of these time-reversal phonon pairs. Two magnons, labeled as $M_1$ and $M_2$ in Fig. \ref{Crystal and Magnetic Structure of MnTe2 Figure 1}(d), appear in the $1.6$ K scan at $2.5$ $\pm$ $0.1$ meV and $3.5$ $\pm$ $0.1$ meV, in agreement with the INS data. These two modes soften and disappear as expected when approaching the ordering temperature (Supplementary Fig. \ref{TempDependenceFull}).

Additional Raman measurements in a magnetic field show that the $2.5$ meV mode is singly degenerate, while the $3.5$ meV mode is triply degenerate and is split into three modes by a magnetic field applied along the low symmetry direction $[0\; 0.7\; 1]$, as shown in Fig. \ref{figure2fittings}(b). We recognize the \textit{irreps} for the magnons as $^1$E$_{\text{g}}$ and T$_{\text{g}}$, respectively. When the field is applied along this low symmetry direction, the $\mathcal{MSG}$ is reduced to $P\bar{1}$, and all degeneracies of the magnon wavefunctions are broken, causing the triply degenerate magnon to split into three separate magnons with the same \textit{irreps} A$_{\text{g}}$. In the Faraday configuration of this experiment where $k\parallel B\parallel$ $[0\; 0.7\; 1]$, all three components of the T$_{\text{g}}$ mode have non-zero Raman response. On the other hand, in the Voigt configuration experiment where $k\parallel (1 0 0)$ and is perpendicular to $B\parallel$ $[0\; 1\; 0.63]$, shown in Fig. \ref{figure2fittings}(d), only two of the three modes are observable due to the form on the Raman tensors of the T$_{\text{g}}$ \textit{irreps} of the $\mathcal{MSG}$ $Pa\bar{3}$ shown in Supplementary Section \ref{Raman-tensors-polarization-temp}.

The INS and magneto-Raman data show highly coherent quasiparticles suggesting that linear spin wave theory (LSWT) can model the spin dispersion correctly.
\ We have extracted a spin model that is consistent with all experimental data to explicitly understand its magnon excitations and further refine our predictions of the magnon topology exhibited in this material. The non-coplanar structure of MnTe$_2$ can be stabilized with an antisymmetric spin-spin nearest-neighbor exchange interaction such as the Dzyaloshinsky-Moriya interaction  (DMI). We then searched for values of the Heisenberg exchange interactions $J_i$ and the DMI $D$ by fitting the LSWT calculations to both the INS spectrum and magnetic field dependence obtained from magneto-Raman measurements. In order to adequately match the observed spin excitations, weaker longer-ranged spin-spin interactions up to the $5^{th}$ nearest neighbors have to be included (Fig. \ref{figure2fittings}a). The overall spin Hamiltonian takes the form:
\begin{equation}
\hspace{-0.2cm}
\label{Spin Hamiltonian}
    \hat{\mathcal{H}}= \sum_{\langle i,j \rangle}[J_1 \hat{\vec{S_i}}\cdot\hat{\vec{S_j}} + \vec{D} \cdot \hat{\vec{S_i}} \times \hat{\vec{S_j}}\;] +\sum_{\langle i,j \rangle _{n=2}}^{5} J_n \; \hat{\vec{S_i}}\cdot\hat{\vec{S_j}} + \\
      g \; \mu_{\mathrm{B}}\sum_{i}\vec{B}\cdot \hat{\vec{S_i}}
\end{equation}

where $J_1$ is the first nearest neighbor (NN) Heisenberg bond, $\vec{D}$ is the NN DMI vector, $J_{n \in \{2,..,5 \}}$ are the $n-$th NN bonds, $g=2$ is the Landé $g$-factor, $\mu_B$ is the Bohr magneton, and $\vec{B}$ is an externally applied magnetic field. 

\begin{table*}[t]
\centering
\begin{tabular}{cccccc}
\toprule
$\boldsymbol{J}_{\boldsymbol{1}}$ & $D_z$ & $D_y$ & $D_x$ & $J_2$ & $J_3^1$ \\
\midrule
$\mathbf{0.5922(139)}$ & 0.09901(281) & 0.08356(151) & $-0.0027456(80)$ &
$-0.11943(340)$ & $-0.05712(169)$ \\
\bottomrule
\end{tabular}

\vspace{0.4cm}

\begin{tabular}{ccccc}
\toprule
$J_3^2$ & $J_4^1$ & $J_4^2$ & $J_5^1$ & $J_5^2$ \\
\midrule
$-0.08659(245)$ & $-0.020076(577)$ & $-0.07905(221)$ &
0.12331(364) & 0.010620(314) \\
\bottomrule
\end{tabular}
\caption{Spin model interaction parameters (in meV). Uncertainties represent standard deviation of an ensemble of parameter sets generated by randomly perturbing the best-fit values within $\pm 5\%$ and retaining only those realizations with a total loss within $15\%$ of the minimum. 
The reported means and standard deviations thus represent the local variability of parameters consistent with the $15 \%$ window around the total loss. 
Bonds, shown in Fig.~\ref{Crystal and Magnetic Structure of MnTe2 Figure 1}(b), of the same length are denoted by superscripts. A full comparison with literature-estimated values for $J_1$, $J_2$, $D$ is presented in Supplementary Table \ref{tab:spinmodels}; roughly, $J_1$ and $J_2$ values are in a good agreement while our DMI strength $|D|$ is around twice that of a recent DFT study. }
\label{tab:spin_model}
\end{table*}

As shown in Fig. \ref{figure2fittings} and Supplementary Figs. \ref{comparisonmap1} \ref{comparisonmap2} \ref{comparisonmap3} \ref{comparisonmap4}, excellent agreement with both INS and magneto-Raman data is obtained via the parameter set in Table \ref{tab:spin_model}. For instance, the spin model is able to accurately fit the  the high-symmetry Brillouin zone map (Fig. \ref{figure2fittings}a), as well as an arbitrarily chosen low-symmetry direction $[H\; 0.6+1.3H\; 0]$, (Fig. \ref{figure2fittings}c).  Furthermore, the spin model is able to reproduce the splitting of the three-fold degenerate $3.5$ meV $\Gamma$-point magnon mode in the Raman spectra, when a magnetic field is applied along the [0 0.7 1] direction (Fig. \ref{figure2fittings}b). This was further confirmed when the magnetic field was applied along [0 1 0.63] (Fig. \ref{figure2fittings}d). 
Further comparisons between the INS measured spectra and the spin model is presented in Supplementary Fig. \ref{comparison_between_observed_vs_calculated}.  The magnitudes of the exchange constants in the spin model suggest dominant nearest neighbor spin-spin interaction, finite DMI stabilizing the non-coplanarity of the magnetic structure, and weaker interactions over the longer bonds.

\begin{figure*}[t]
    \hspace{0cm}
\includegraphics[width=1.0\linewidth]{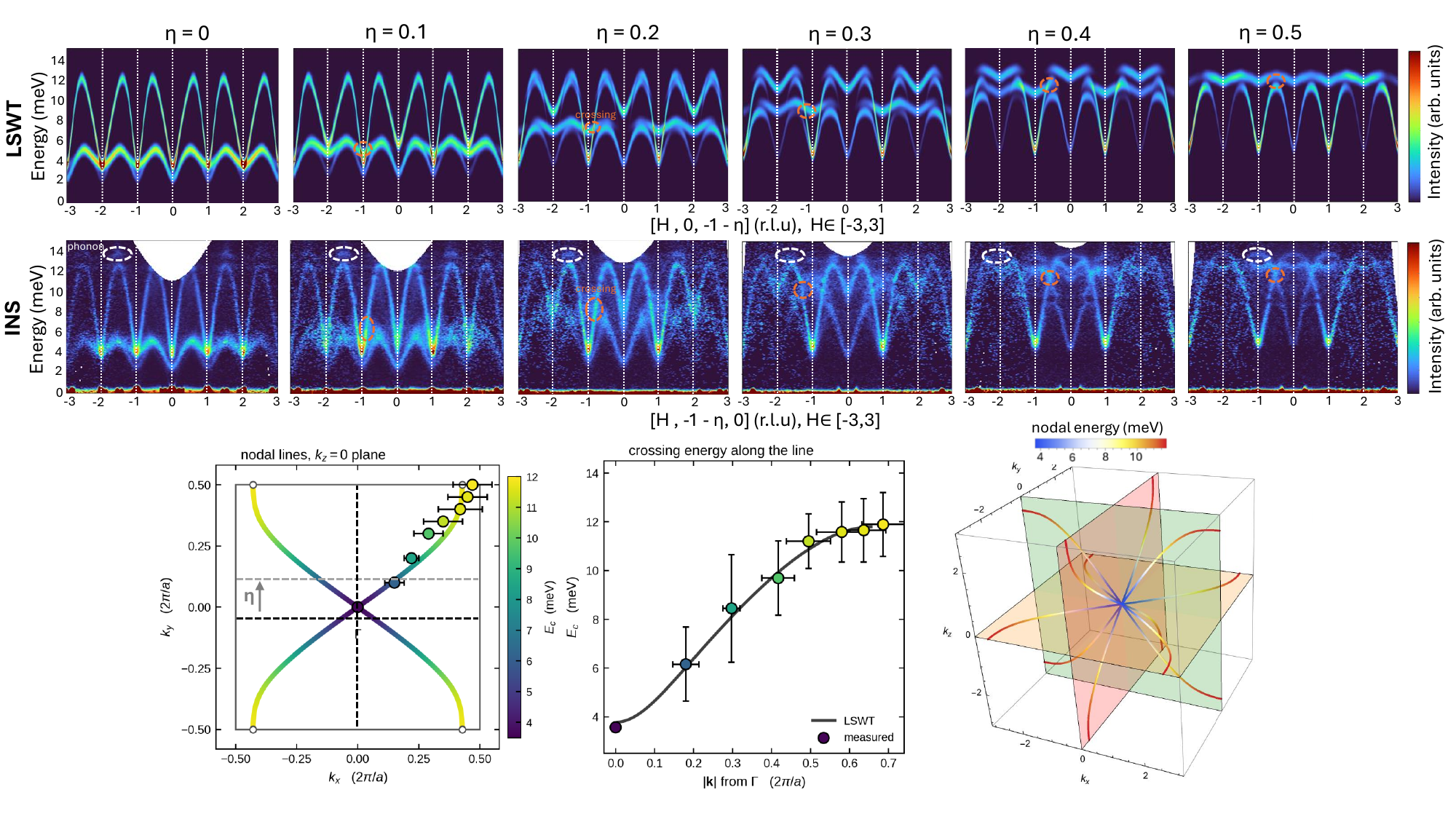}
 \caption{\textbf{Extracting the nodal lines from INS.} First row: calculated INS intensity on the $[H,0,-1-\eta]$ cuts. Different values of $\eta$ represent an intersection through through the nodal line, as marked by the orange dashed circles. Second row: measured INS intensity on the $[H,0,-1-\eta]$ cuts. The dashed orange crossings are used to construct the structure of the nodal line as plotted in the third row. Direct comparison of the first two rows discriminates the $\sim 13$ meV phonon modes marked by the white ellipses in the second row. Third row: comparison between the extracted crossing points and the calculated nodal line. The rest of the nodal line is constructable through the symmetry operations within the plane. The parameter $\eta$ marks the cuts used in the first two rows.
}

    \label{mnte2_Extract_Nodal_Line}
\end{figure*}

Magnon band topology calculations were performed using the obtained spin Hamiltonian with an additional magnetic field along a low-symmetry direction, which we take to be along [0 1 6]. First, we calculate the inversion eigenvalues of the magnon eigenfunctions at the eight high-symmetry momenta, as presented in the lower part of Fig. \ref{figure3weylpointsinfield}(a). From this, the calculated $\mathcal{SI}s$ were confirmed to belong to the values predicted by band representations \cite{karaki2024highthroughputsearchtopologicalmagnon}. In particular, we find that the second and third magnon bands cross at some generic momenta in the BZ, and therefore we can interpret the middle gap $\mathcal{SI}$ value $1112$ as following. First, the $\mathbb{Z}_{2,i}$ index indicates that the Chern number of the combined upper two bands takes a value of $1 \; mod \; 2$ on the high symmetry planes $k_i=0,\pi$ for $i=x,y,z$. This is confirmed explicitly in our model as seen in Fig. \ref{figure3weylpointsinfield} (b,c), calculated using the numerical technique developed in Ref. \cite{ChernNumberCalculationFukui_2005}. Second, the $\mathbb{Z}_4$ index indicates a difference of $2$ between the Chern numbers on the $k_z=0,\pi$ planes. This is an indication of a gap closing and an existence of an even number of Weyl magnons in between these two high symmetry planes. Additionally, the chiralities of these Weyl magnons are such that they give rise to the difference of the Chern numbers on these two planes.

\begin{figure}[H]
    \centering
    \includegraphics[scale=0.4]{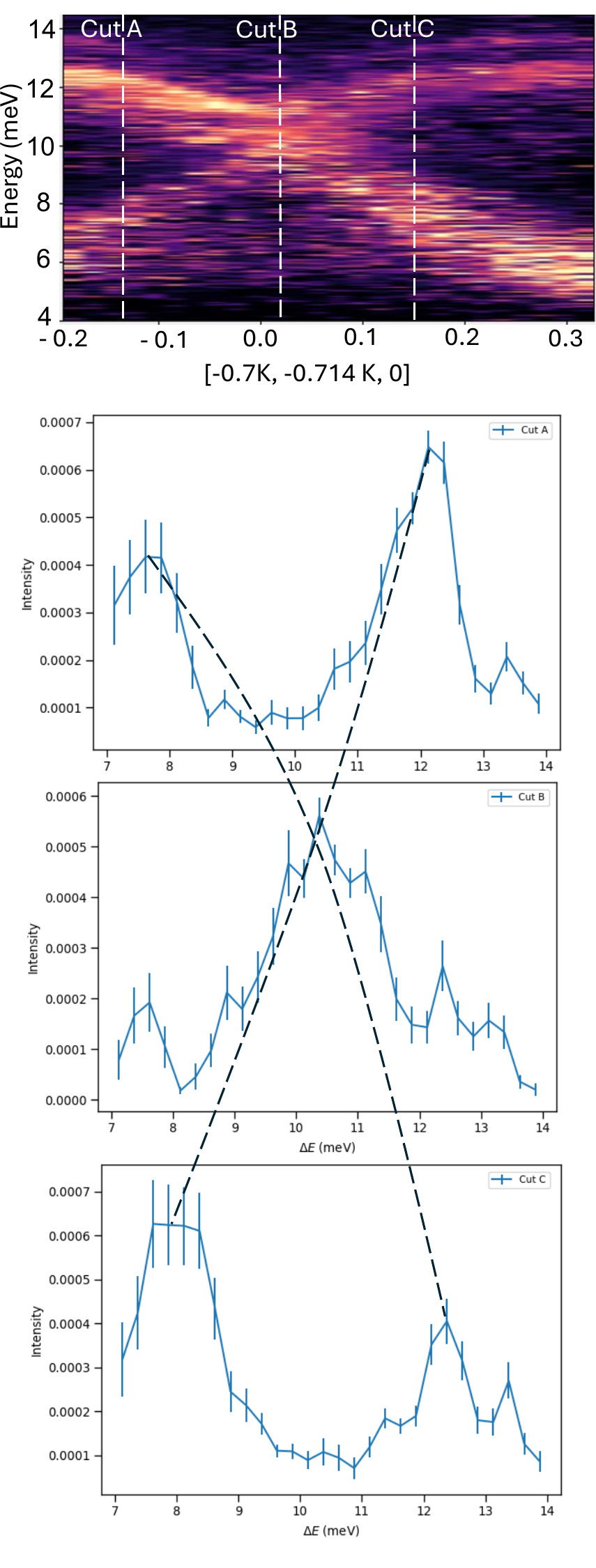}
    \caption{\textbf{Probing magnon nodal line crossing.} (a) The INS extracted magnon dispersion along the
in-plane direction $q_\perp\parallel(\bar{0}.700,\bar{0}.714,0)$ passing through the point $\mathbf{Q}=(1.26,\bar{1}.33,0)$ which lies on the nodal line with crossing energy $E_c=9.96$~meV. Data are collected with an integration range of the in-plane perpendicular momentum $[\bar{0}.714H,0.7001H,0]$ of  $H\in [-1.85,-1.81]$ and $[0,0,L]$ $ \in [-0.15,0.15]$. Three 1D energy cuts before (cut A), at (cut B) and after (cut C) are analyzed in panels (b-d) to show a more quantitative map of the crossing of the two magnon branches, one of them being more intense than the second. An extra integrations range of $[\bar{0}.700,\bar{0}.714,0] \in \{ [-0.15,-0.11], [0.0,0.03], [0.13,0.71] \}$ is adopted for panels (b,c,d), respectively. Dashed lines trace the intensity peaks of cut (A) as they cross (Cut B) and split apart again in cut (C). The different color map used in panel (a) is for visual clarity.  }
    \label{1D_cuts_nodal_line}
\end{figure}

Throughout, we focus on the second energy band gap for two reasons: first, this topological band gap\cite{doi:10.1126/sciadv.ade7731,karaki2024highthroughputsearchtopologicalmagnon} occurs for all possible spin models, and second, it is also larger than the first and third band gaps. The subgroup $P\bar{1}$ is special in the sense that regardless of the order of the magnon band \textit{irreps} of $Pa\bar{3}$, they will always decompose into \textit{irreps} of $P\bar{1}$ with a topological  gap between the second and third magnon bands. The upper panel of Fig. \ref{figure3weylpointsinfield}(a) shows a schematic representation of the magnon band \textit{irreps} which, upon applying the appropriate perturbation to break into $P\bar{1}$, are decomposed into \textit{irreps} of $P\bar{1}$ such that the middle gap is always topological as discussed above. The calculated $\mathcal{SI}$s of the induced band gaps using our spin model are shown in the lower panel of Fig. \ref{figure3weylpointsinfield}(a). Scanning the BZ between the second and third bands, we identified two Weyl crossings at generic momenta $\boldsymbol{k}=(0.4964, 0.0031, 0.4333),(-0.04969, 0.2142, 0.1518)$ in units of $2\pi /a$, where $a$ is the lattice constant, at energies $11.75$ meV and $7.48$ meV, in a magnetic field of $14$ T along $[0 \; 1 \; 6]$. Their topological charges, calculated as the Chern number on a small sphere enclosing the crossing point are $+1,+1$, and are consistent with the Chern numbers on the high-symmetry planes $k_i=0,\pi$ for $i=x,y,z$. Inversion symmetry enforces their opposite-chirality partners to lie at the opposite momenta. Fig. \ref{figure3weylpointsinfield} (d) shows the position of the two Weyl pairs. In the absence of a magnetic field, the $\mathcal{MSG}$ contains three glide symmetries $\{m_{100}|\frac{1}{2} \frac{1}{2} 0 \}$, $\{m_{010}|0 \frac{1}{2} \frac{1}{2} \}$, and $\{m_{001}|\frac{1}{2} 0 \frac{1}{2}\}$ that protect magnon nodal lines\cite{Fang_2016} on the glide-symmetric planes $k_i=0$ for $i=x,y,z$ (more details in Supplementary Section \ref{topo-of-middle-band-gap}), as validated by the spin wave calculation and confirmed with the INS data as presented in Fig. \ref{mnte2_Extract_Nodal_Line}. The X-shaped crossing for a point on the nodal line is shown in Fig. \ref{1D_cuts_nodal_line}. The topology of these nodal lines is confirmed in terms of a nonzero Berry phase of $\pi$ ($ mod \; 2\pi$) of the upper two bands on a closed contour that encircles the magnon nodal lines. By tracing the locations of the Weyl magnons (shown in Fig. \ref{figure3weylpointsinfield} (d)) as the magnetic field is turned off, we find that Weyl pairs move in the BZ towards the high-symmetry planes until they return and become part of the nodal lines at zero field, see Supplementary Fig. \ref{weyl-emergence}. In other words, as the magnetic field is turned on, the zero-field nodal lines gap out and the two Weyl magnon pairs emerge.

\begin{figure*}[t]
 \hspace{-0cm}
 \includegraphics[scale=0.68]{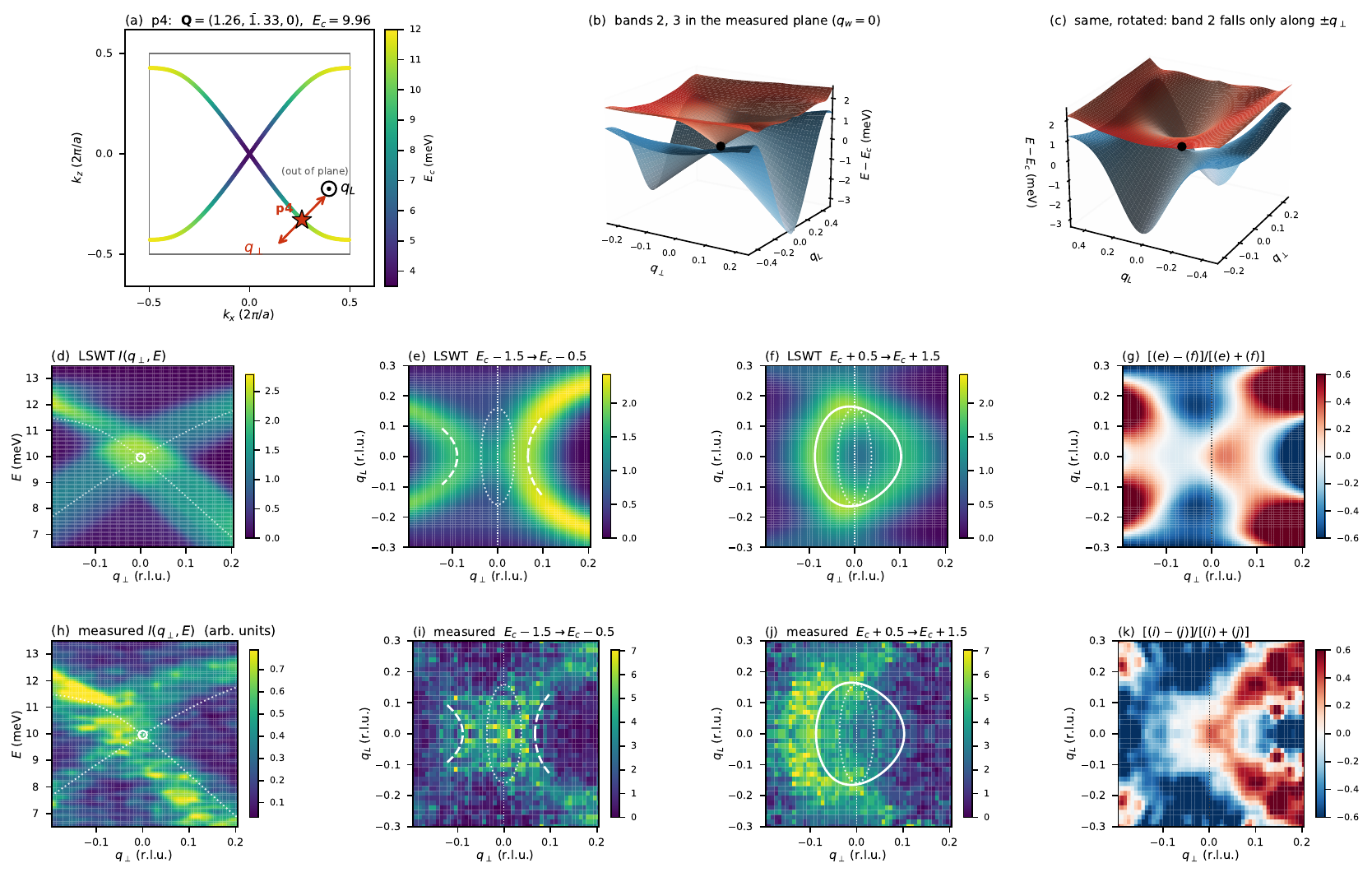}
 \caption{\textbf{A point on the nodal line with the associated INS intensity winding.} (a)~Position of a chosen point on the nodal line in the $k_y=0$ plane of the magnetic Brillouin zone. The
coloured lines trace the loci where bands~2 and~3 become degenerate, coloured by
the degeneracy energy $E_c$; the star marks the chosen point at
$\mathbf{Q}=(1.26,\bar{1}.33,0)$, $E_c=9.96$~meV. The double arrow gives the
in-plane direction $q_\perp\parallel(\bar{0}.700,\bar{0}.714,0)$ across which the
touching is measured; the second momentum axis of the slices,
$q_L\parallel(0,0,1)$, is perpendicular to the plane shown.
(b,c)~Bands~2 (blue) and~3 (red) in the $(q_\perp,q_L)$ plane at $q_w=0$, with
energy as the vertical axis, from two viewing directions; and the black sphere marks the degeneracy.  The
touching is conical only along $\pm q_\perp$: band~3 forms a bowl that rises in
every direction, whereas band~2 is a saddle that descends steeply along
$\pm q_\perp$ ($-3.6$~meV at $q_\perp=0.24$) but falls by only $0.09$~meV along
$q_L$ before curving back up. The touching
point is tilted along the nodal line, so that
the crossing is of type~II in that direction.
(d--g)~Linear spin-wave prediction and (h--k)~the corresponding measurement, both
in the frame of~(a).
(d,h)~Intensity in the $(q_\perp,E)$ plane, integrated over
$|q_L|\le0.15$ and $|q_w|\le0.06$~r.l.u.; the open circle marks $E_c$ and the
dotted curves are the calculated band-2 and band-3 dispersions at $q_L=0$.
(e,f) and~(i,j)~Constant-energy maps in the $(q_\perp,q_L)$ plane, integrated over
$[E_c-1.5, E_c-0.5]$~meV (below) and $[E_c+0.5, E_c+1.5]$~meV (above) and over a
$w$~slab of $\pm0.06$~r.l.u.  The white curves are calculated contours of
constant branch energy: solid, $E_3=E_c+1$~meV, which closes about the touching
point; dashed, $E_2=E_c-1$~meV, which does not close because band~2 turns over
along $q_L$; dotted, the cone-matched loop of semi-axes
$(a_\perp,a_L)=(0.037,0.159)$~r.l.u.\ used to extract the intensity winding.
(g,k)~Normalised difference of the two windows,
$(I_\downarrow-I_\uparrow)/(I_\downarrow+I_\uparrow)$, after subtracting a
flat background from each measured map.  The antisymmetry about $q_\perp=0$ is
the intensity winding of the two crossing branches.}
 \label{figure5intensitywinding}
 \end{figure*}

A direct experimental signature of the magnon nodal lines at zero field is the way in which the INS intensity changes around a loop in momentum space for energy ranges above and below it as established in Ref. \cite{shivam2017neutronscatteringsignaturesmagnon} for arbitrary magnon two-band touching points. This winding behavior is directly related to the non-trivial topology of the magnon wavefunction near the nodal crossing, and has been verified experimentally for the case of Dirac magnons \cite{ThermalEvolutionOfMagnonsinCrBr3PhysRevLett.129.127201,GadoliniumMcClartyPhysRevLett.128.097201} and Dirac nodal lines\cite{CoTiO3McClartyNatComm.10.1038/s41467-021-23851-0}. We demonstrate this behavior here, as a result of the nonzero Berry phase around the nodal lines, by focusing on a point that belongs to one of the nodal lines at zero field, the $\mathbf{Q}=(1.26,\bar{1}.33,0)$ point at a crossing energy $E_c=9.96$~meV (see Fig. \ref{figure5intensitywinding} (a)), and analyze the INS intensity on two constant energy intervals, one below and one above the energy of the nodal crossing. As discussed in Fig. \ref{figure5intensitywinding}, the INS intensity follows a modulation while winding around the nodal 'cone' for an energy range above the node energy, and follows an opposite winding for a lower energy range (Fig \ref{figure5intensitywinding}b). We also compare the measured modulation to the LSW calculated one and show a good agreement between the two.

\emph{Experimental Signatures Predictions. }Weyl magnons act as sources and sinks of magnon Berry curvature (see Supplementary Fig. \ref{fig:sm-monopolefig} and Supplementary Section \ref{berry-curvature-and-flux} for more details); therefore, a direct experimentally-accessible consequence for the existence of Weyl magnons at finite field is the magnon thermal Hall conductivity, allowed by the field-breaking of the $T_h$ cubic symmetry (See Supplementary Section \ref{sec:thermal-hall}), calculated using our spin model to be of order $\kappa_{xy}\sim 10^{-3} \;WK^{-1}m^{-1}$ . Given the existence of phonons within the same energy range of magnons, as shown in the high-temperature INS measurements of Supplementary Section \ref{phonon-high-temp-section} and as indicated from Fig. \ref{mnte2_Extract_Nodal_Line}, we anticipate that our pure-magnon thermal Hall predictions to be modified towards higher values due to the possible magnon-phonon hybridization (as hinted from the $\sim 8$ meV avoided-crossings on the $X-M$ path in Fig. \ref{mnte2_Extract_Nodal_Line}'s $\eta=0.5$ cut) that can act as extra hotspots of Berry curvature. Additionally, according to the bulk-boundary correspondence, the non-trivial topology of the bulk is associated with the presence of surface modes, which we analyze in detail in Supplementary Section \ref{surface-modes}.

We have established that the non-coplanar antiferromagnet MnTe$_2$ is a material host for topological nodal line magnons in zero field and Weyl magnons at finite field. Through a combination of INS, magneto-Raman measurements, and theoretical calculations we probed the magnon band structure of MnTe$_2$, determined the microscopic spin Hamiltonian, confirmed the existence of topological magnon nodal lines enforced by the $\mathcal{MSG}$ symmetries, and analyzed the emergence of Weyl magnons induced by an external magnetic field. This discovery not only establishes a concrete bosonic analogue of Weyl fermions but also provides a promising platform for exploring tunable topological excitations and associated transport phenomena in magnetic systems. Our results serve as the foundation for future investigations of direct in-field INS imaging of Weyl magnons, their associated bulk thermal transport \cite{ThermalHallOnose_2010} and surface magnons, and possible device applications based on topological spin transport \cite{Chumak2015}. This realization of a tunable Weyl magnon phase highlights the rich interplay between magnetic order and band topology in non-coplanar antiferromagnets and demonstrates the power of combining the symmetry based topological analysis with appropriate experiments in the discovery of new topological magnon materials.

\textit{Acknowledgements}--
This work was primarily funded by the National Science Foundation MRSEC program under grant DMR-2011876 to the Center for Emergent Materials at Ohio State. The identification of any commercial product or trade name does not imply
endorsement or recommendation by the National Institute of Standards and Technology.  This research used resources at the Spallation Neutron Source, a DOE Office of Science User Facility operated by the Oak Ridge National Laboratory. The beam time was allocated to SEQUOIA (BL-17) on proposal number IPTS-32637.1.

\bibliographystyle{apsrev4-2}
\bibliography{main}

\appendix
\section{Appendix: Experimental Details}
The inelastic neutron scattering experiments were performed on the Fine-Resolution Fermi Chopper Spectrometer \cite{GRANROTH20061104} at the Spallation Neutron Source, Oak Ridge National Laboratory. A 1.7 g single crystal of pure MnTe\textsubscript{2} was used with incident neutron energies of E\textsubscript{i} = 22 meV.  The chopper configuration consisted of the T-zero chopper operating at 60 Hz and the instruments standard fine resolution Fermi chopper operating at 240 Hz.  This configuration has a calculated full width at half maximum (FWHM) energy resolution 0.55 meV for elasticly scattered neutrons and 0.41 meV at 10 meV energy transfer.  Measurements were performed at  $T = 5$~K.\\

The crystal was aligned prior to the experiment using a backscattering Laue X-ray diffractometer.   The crystal was aligned in the (HK0) scattering plane at the neutron spectrometer.  Alignment scans were performed with 81.81 meV incident energy neutrons with the sample at $T=5$~K.  The full width at half maximum intensity of the nuclear peak rocking curves was measured as 0.78 degrees for the (4 0 0) and 0.88 degrees for the (0 -4 0) reflection respectively.  The measured energy resolution of elastically scattered neutrons using the nuclear incoherent scattering of the sample was found to be 0.61 meV FWHM.\\

The crystal was mounted in an aluminum sample can and sealed under an atmosphere of He gas. A neutron-absorbing Cd foil was placed at the bottom of the sample can and in the vicinity of the sample mounting hardware to reduce the neutron scattering background from the sample holder. In order to access a volume of reciprocal space with the instruments highly pixelated detector, the sample was rotated about its vertical axis over a range of 140 degrees with a step size of 0.5 degrees.  Individual angles of this rotation were measured for between 0.47 and 1.67 Coulombs of charge on the spallation neutron source target.  This corresponds to between 6 and 21 minutes of data collection for each angle measured.  The data were histogrammed using the SHIVER software and were folded into the first Brillouin zone of the simple cubic $Pa\bar{3}$ space group of MnTe$_2$ to improve counting statistics. \\



Magneto-Raman scattering at NIST was done using a $632$ nm  He-Ne laser with a triple grating spectrometer which has a resolution around 0.4 cm$^{-1}$. Experiments in both Faraday and Voigt geometries ($k\parallel B$ and $k\perp B$, respectively) were performed. The sample is mounted in a cryostat which cools down the sample from room temperature to around 1.6K with the sample in a He exchange gas environment. This system is also able to apply an external magnetic field up to 9 T. Polarization optics were used in both the incoming and outgoing beams, allowing linear polarization analysis. The parallel polarizers configuration results are acquired by rotating a half-wave plate in front of the sample. Laser power was maintained at 300 $\mu$W for the $B\parallel (0\ 0.7\ 1)$ and at 150 $\mu$W for the $B\parallel (0\ 1\ 0.63)$ experiments to avoid any possible damage. 

\nocite{GoldbergerPdSe2,MnTe2NoncollinearMagnetismPhysRevB.56.14013,lu2018magnonbandtopologyspinorbital,doi:10.1126/sciadv.ade7731,karaki2024highthroughputsearchtopologicalmagnon,DFT-Hohenberg,DFT-Kohn,VASP-1996,VASP-1999,VASP-2006,giannozzi2009quantum,PBEsol,DFTU,Phonopy2023,RevPhonopy2023}

\clearpage
\onecolumngrid

\setcounter{secnumdepth}{3}

\renewcommand{\thesection}{\Roman{section}}
\renewcommand{\thesubsection}{\Alph{subsection}}
\renewcommand{\thesubsubsection}{\arabic{subsubsection}}

\begin{center}

{\Large\bfseries \textit{Supplemental Materials}}\\[1.0em]

{\large\bfseries
Towards The First Realization of Weyl Magnons in the Non-Coplanar Antiferromagnet MnTe$_2$
}\\[1.0em]

{\normalsize Ahmed E. Fahmy et al.}

\end{center}

\vspace{1.5em}


\setcounter{section}{0}
\renewcommand{\thesection}{\Roman{section}}
\renewcommand{\thesubsection}{\Alph{subsection}}
\renewcommand{\thesubsubsection}{\arabic{subsubsection}}

\renewcommand{\figurename}{Supplementary Figure}
\setcounter{figure}{0}
\renewcommand{\thefigure}{\arabic{figure}}

\renewcommand{\tablename}{Supplementary Table}
\setcounter{table}{0}
\renewcommand{\thetable}{\arabic{table}}


\renewcommand{\theHfigure}{SI.figure.\arabic{figure}}
\renewcommand{\theHtable}{SI.table.\arabic{table}}
\renewcommand{\theHsection}{SI.section.\arabic{section}}
\renewcommand{\theHsubsection}{SI.subsection.\arabic{section}.\arabic{subsection}}
\renewcommand{\theHsubsubsection}{SI.subsubsection.\arabic{section}.\arabic{subsection}.\arabic{subsubsection}}

\setcounter{section}{0}
\renewcommand{\thesection}{\Roman{section}}

\tableofcontents

\section{Crystal Growth}
\label{crsytal-growth}
\subsection{Synthesis}

 Single crystals of MnTe\textsubscript{2} were grown from a Bridgman type growth using Te self-flux. Mn powder (99.95\%, Thermo Scientific) and Te lump (99.9999\%, Thermo Scientific) were mixed together in a 1:4 molar ratio and ground to a coarse powder at a typical total scale of 20~g. This mixture was added to a quartz ampoule under an Ar atmosphere and sealed under \(<\)80 mTorr. To maximize single crystal growth, these ampoules were closed with a tapered end rather than a rounded end, and the bottom of the tube was graphitized with isopropyl alcohol. The tube was hung in a vertical single-zone furnace by the procedure outlined in \cite{GoldbergerPdSe2}. The ampoule was heated to 850 °C over 24 hours and held at temperature for 50 hours to allow homogenization. The tube was then dropped at 5.79 \(\mu\)L/hr or the equivalent of 0.33 mm/hr over 330 hours before allowing to cool to room temperature over 12 hours. Single crystals of MnTe\textsubscript{2} (Supplementary Fig. \ref{CrystalGrowth}a) were extracted from the flux through the use of a razor blade.

\subsection{Structural Characterization}

 Phase purity of the grown MnTe\textsubscript{2} crystals was confirmed by grinding grown crystals into a powder for analysis using powder X-ray diffraction. Powder X-ray diffraction measurements were performed on a Bruker D8 Advance powder X-ray diffractometer using Cu K\textsubscript{\(\alpha\)1} radiation that was generated at 40 kV and 40 mA.

\begin{figure}[H]
    \centering
    \includegraphics[width=0.4\linewidth]{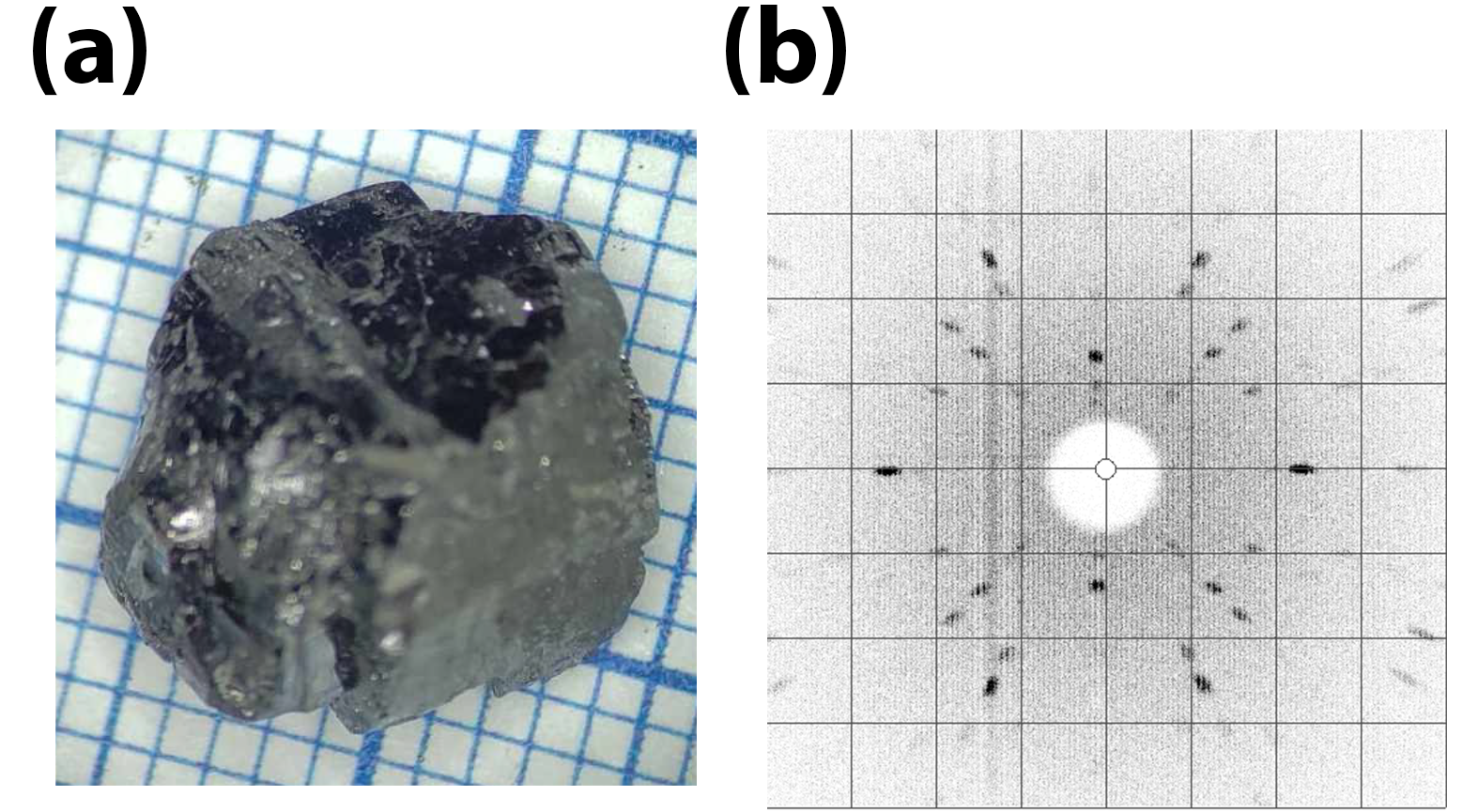}
    \caption{(a) Optical image of MnTe\textsubscript{2} single crystal on mm-grid paper.(b) Representative backscattering Laue diffraction pattern along the 1 0 0 direction indicating single crystallinity.}
    \label{CrystalGrowth}
\end{figure}

 Single crystallinity across the grown crystals was confirmed by Laue backscatter diffraction (Supplementary Fig. \ref{CrystalGrowth}b). Laue diffraction measurements were performed on a Multiwire MWL 120 diffractometer using a white light Cu X-ray source generated at 20 kV and -9.5 mA.
 
\section{Magnetic Measurements}
\label{magnetic-measurements}
\begin{figure}[H]
     \centering
     \includegraphics[width=0.5\linewidth]{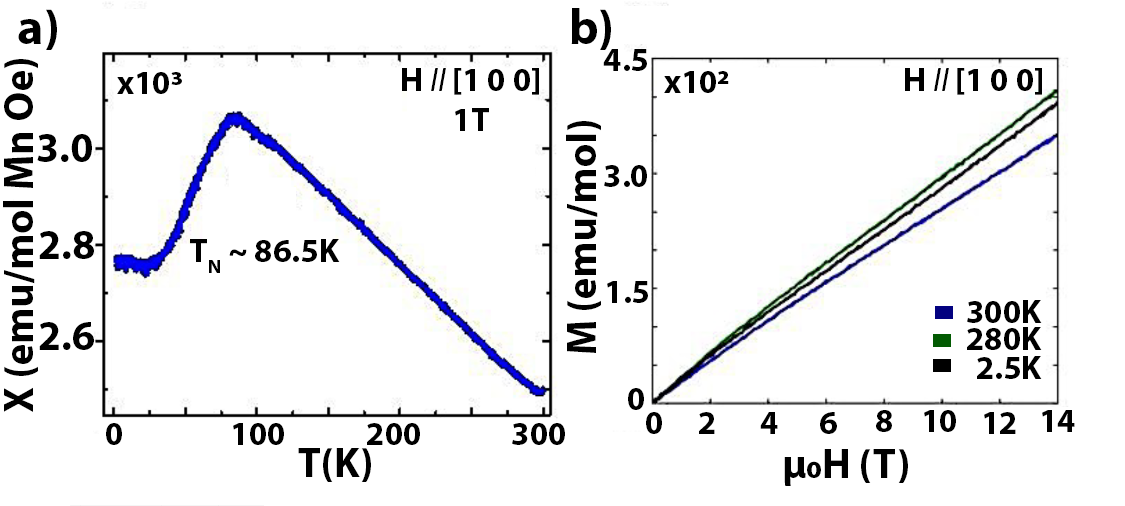}
     \caption{(a) Measurement of single crystal magnetic susceptibility vs. temperature with with a $1$ T applied field along [1 0 0]. (b) Measurement of magnetization v.s. field below and above the magnetic transition.}
     \label{magnetizationandsusceptibility}
 \end{figure}

Magnetic susceptibility measurements were obtained using a Quantum Design MPMS3 superconducting quantum interference device (SQUID). Ground crystals were added to an inverted gelatin capsule that was affixed together by Kapton tape. The gelatin capsule was then inserted into a straw and positioned for measurement. Temperature dependent magnetic susceptibility (\(\chi\) vs. T) was collected upon warming from $2$ K (Supplementary Fig. \ref{magnetizationandsusceptibility}a); zero-field cooled data acquisition was done first using a 500 Oe measurement field and then field cooled data acquisition followed by cooling and measuring with the same $500$ Oe field. Magnetization vs. Field (M vs. H) measurements were performed at $2$ K after zero field cooling before the field was swept from 0 to $\pm$ $4T$ and which trace was run twice (Supplementary Fig. \ref{magnetizationandsusceptibility}b).

\section{Linear Spin Wave Theory}

The crystal structure of MnTe$_2$ belongs to the space group \textit{Pa$\bar{3}$} (205) which has four Mn sublattices at the $\mathcal{WP}$ 4a:
\begin{equation}
    \begin{aligned}
        & r_A = (0, 0, 0) \\
        & r_B = (a/2,0,a/2) \\
        & r_C = (0, a/2, a/2) \\
        & r_D = (a/2, a/2, 0)
    \end{aligned}
\end{equation}
where Cartesian coordinates are used and $a$ is the (cubic) lattice constant. In the magnetic phase below $T_N \sim 87$ K, the Mn atoms become magnetized giving rise to the non-coplanar magnetic order \cite{MnTe2NoncollinearMagnetismPhysRevB.56.14013}:
\begin{equation}
    \begin{aligned}
        & \langle \boldsymbol{S}_{A} \rangle \propto (+1,+1,+1) \\
        &\langle \boldsymbol{S}_{B} \rangle \propto (-1,-1,+1)\\
        &\langle \boldsymbol{S}_{C} \rangle \propto (-1, +1,-1)\\
        &\langle \boldsymbol{S}_{D} \rangle \propto (+1, -1, -1).
    \end{aligned}
\end{equation}

We build our spin model based on the interaction bonds defined between pairs of sites as following:\\

\begin{equation}
\begin{aligned}
    & \mathbf{J_1} : r_B \iff r_B + (a/2,0,-a/2) \\
    & \mathbf{J_2} : r_A \iff r_A + (a,0,0) \\
    & \mathbf{J_3^1} : r_B \iff r_B + (a/2, -a, -a/2) \\
    & \mathbf{J_3^2} : r_B \iff r_B + (a/2, 1, -a/2) \\
    & \mathbf{J_4^1} : r_A \iff r_A + (a, 0, -a) \\
    & \mathbf{J_4^2} : r_A \iff r_A + (a, a, 0) \\
    & \mathbf{J_5^1} : r_B \iff r_B + (a/2, 0, -3a/2) \\
    & \mathbf{J_5^2} : r_B \iff r_B + (3a/2, 0, -a/2) \\
\end{aligned}
\end{equation}
where $\mathbf{J_2}$, $\mathbf{J_4^1}$ and $\mathbf{J_4^2}$ are constrained by inversion $\{-1|0\}$ and therefore take the general symmetric form:
\begin{equation}
    \mathbf{J_{i=2,4}} =
\begin{pmatrix}
  J_i^{xx} & \Gamma_i^{xy} & \Gamma_i^{xz}\\
  \Gamma_i^{xy} & J_i^{yy} & \Gamma_i^{yz}\\
  \Gamma_i^{xz} & \Gamma_i^{yz} & J_i^{zz}
\end{pmatrix}
\end{equation}
and the rest of bonds are not symmetry-constrained, so therefore take the general form:
\begin{equation}
    \mathbf{J_{i\neq2,4}} = 
\begin{pmatrix}
  J_i^{xx} & D_i^z + \Gamma_i^{xy} & -D_i^y + \Gamma_i^{xz}\\
  -D_i^z + \Gamma_i^{xy} & J_i^{yy} & D_i^x + \Gamma_i^{yz}\\
  D_i^y + \Gamma_i^{xz} & -D_i^x + \Gamma_i^{yz} & J_i^{zz}
\end{pmatrix}
\end{equation}

We calculate the magnon band structure using the equation of motion (EOM) approach \cite{lu2018magnonbandtopologyspinorbital} in which the low-energy dynamics is captured by the small deviations of spins from their ordered directions :
\begin{equation}
    \mathbf{s}_i \equiv \mathbf{S}_i-\left\langle\mathbf{S}_i\right\rangle \Rightarrow \mathbf{s}_i=\left(s_i^x, s_i^y, 0\right) \label{eq:LSW EOM Expansion}
\end{equation}
where ($S^x$, $S^y$) are the transverse spin components in its local frame that is locally rotated such that the z-axis is aligned along the polarization direction $\langle \boldsymbol{S}_i \rangle = S_i \hat{z}_i = (0,0,S_i)$, and $\left|\mathbf{s}_i\right| \ll \bar{S}_i $. By expanding the Hamiltonian in terms of the deviations, then considering their Heisenberg EOM, we reach a dynamical EOM that governs the spinwave dynamics:
\begin{equation}
    -\mathrm{i} \frac{\mathrm{d} s_i^\alpha}{\mathrm{d} t} = \sum_{j, \beta}\left(\mathbf{M} \cdot \mathbf{R}\right)_{i \alpha, j \beta} s_j^\beta  \label{eq:LSW EOM R Matrix}
\end{equation}
where $\mathbf{M}_{i \alpha, j \beta} = \delta_{i, j}\left(\sigma_y\right)^{\alpha, \beta}$ with $\sigma_y$ is y-Pauli matrix and $\mathbf{R}$ is a $2N \times 2N$, where $N$ is the number of spins in the primitive magnetic unit cell, is a semi-positive definite hermitian matrix that describes the spin-spin interactions and is referred to as the ``magnon'' Hamiltonian. The spectrum is then calculated as the eigenvalues of $\mathbf{M}\cdot\mathbf{R}$. Using the constructed spin model in Table 1, we obtain the magnon Hamiltonian $\mathbf{R}(\boldsymbol{k})$, diagonalize $\mathbf{M}\cdot\mathbf{R}(\boldsymbol{k})$ and find the magnon spectrum  as shown in Fig. 2 and the magnon wavefunctions which are used to calculate the \textit{irreps} of magnon bands at the high-symmetry momenta.

To directly compare with the experimental data, as shown in Supplementary  Figs. \ref{comparisonmap1} \ref{comparisonmap2} \ref{comparisonmap3} \ref{comparisonmap4}, we compute the neutron scattering cross section, which is given by
\begin{equation}
\frac{\mathrm{d}^2 \sigma}{\mathrm{d}\Omega\, \mathrm{d}E}
\propto
\sum_{\alpha\beta}
\left(
\delta_{\alpha\beta}
-
\frac{Q_\alpha Q_\beta}{Q^2}
\right)
S^{\alpha\beta}(\mathbf{Q},E),
\label{eq:cross_section}
\end{equation}
where $\alpha,\beta = x,y,z$ denote cartesian components, $Q_\alpha$ is the $\alpha$-component of the momentum transfer $\mathbf{Q}$, and $S^{\alpha\beta}(\mathbf{Q},E)$ is the dynamical spin--spin correlation function.

The latter is defined as
\begin{equation}
S^{\alpha\beta}(\mathbf{Q},E)
=
\frac{1}{N}
\sum_{i j}
\mathrm{e}^{\mathrm{i}\mathbf{Q}\cdot(\mathbf{r}_i-\mathbf{r}_j)}
\int_{-\infty}^{\infty}
\langle S_i^{\alpha}\, S_j^{\beta}(t) \rangle\,
\mathrm{e}^{-\mathrm{i}Et}\,
\mathrm{d}t,
\label{eq:structure_factor}
\end{equation}
where $N$ is the total number of spins, $\mathbf{r}_i$ denotes the position of site $i$, and $S_i^{\alpha}$ is the $\alpha$-component of the effective spin operator $S_i$.

\section{Comparison of the spin model to literature estimates}

A Comparison of the spin model to literature estimates is presented in Supplementary Table \ref{tab:spinmodels}.

\begin{table}[h]
\caption{\label{tab:spinmodels}%
\textbf{Exchange parameters reported for pyrite MnTe$_2$ ($Pa\bar{3}$, Mn$^{2+}$, $S=5/2$) in literature.}
All entries are expressed in the convention of Eq.~(2),
$\hat{\mathcal{H}}=\sum_{\langle i,j\rangle}\bigl[J_{1}\hat{\bm{S}}_{i}\!\cdot\!\hat{\bm{S}}_{j}
+\bm{D}\!\cdot\!(\hat{\bm{S}}_{i}\!\times\!\hat{\bm{S}}_{j})\bigr]
+\sum_{n=2}^{5}\sum_{\langle i,j\rangle_{n}}J_{n}\hat{\bm{S}}_{i}\!\cdot\!\hat{\bm{S}}_{j}$,
in which every bond is counted once and $J>0$ denotes antiferromagnetic coupling.
Refs.~\onlinecite{Sawaoka1966,Lin1968,Okada1977,vanKalkeren1978,Okada1980} instead write
$\hat{\mathcal{H}}=-2\sum_{i<j}J_{ij}\hat{\bm{S}}_{i}\!\cdot\!\hat{\bm{S}}_{j}$ and quote $J$ in
kelvin, so their results enter here as $J_{n}=-2J_{n}^{\mathrm{pub}}$; the published numbers are
reproduced in the final column.\protect\footnotemark[1]
All exchange parameters and $|\bm{D}|$ are given in meV, using
$1\,\mathrm{K}=0.086173$~meV.
Uncertainties are those stated by the original references; entries given without one were published
as bare values. $\chi$ referes to magnetic susceptibility measurements, and $P$ refers to pressure-dependent measurements.
}
\begin{ruledtabular}
\setlength{\tabcolsep}{2pt}
\footnotesize
\begin{tabular}{l l
                S[table-format=+1.4(3)]
                S[table-format=+1.5(2)]
                S[table-format=1.4(2)]
                S[table-format=1.3(2)]
                l}
Reference & Method
 & {$J_{1}$} & {$J_{2}$} & {$|\bm{D}|$} & {$|\bm{D}|/J_{1}$} & {Published (K)} \\
\colrule
Sawaoka \textit{et al.}\ (1966)\cite{Sawaoka1966}
 & $\chi(T)$, $T_{N}(P)$; molecular field analysis
 & 1.0341 & 0.2585 & {\textemdash} & {\textemdash} & $-6.0$, $-1.5$ \\
Lin and Hacker (1968)\footnotemark[2]\cite{Lin1968}
 & $\chi(T)$; Curie--Weiss $\theta$ and $T_N$
 & 1.1099 & 0.2068 & {\textemdash} & {\textemdash} & $-6.44$, $-1.20$ \\
Okada and Miyadai (1977)\footnotemark[3]\cite{Okada1977}
 & $\chi(T)$; $\chi(T_{N})$ sum rule, high-$T$ fit
 & 1.224(86) & -0.224(86) & {\textemdash} & {\textemdash} & $-7.1(5)$, $+1.3(5)$ \\
van Kalkeren \textit{et al.}\ (1978)\footnotemark[4]\cite{vanKalkeren1978}
 & Effective-electron superexchange
 & 0.5170 & -0.0517 & {\textemdash} & {\textemdash} & $-3.0$, $+0.3$ \\
Okada (1980)\footnotemark[5]\cite{Okada1980}
 & $\chi(T)$ of dilute Mn clusters
 & 0.7756 & 0.1723 & {\textemdash} & {\textemdash} & $-4.5$, $-1$, $-1$ \\
Heinil\"a and Oja (1994)\footnotemark[6]\cite{Heinila1994}
 & Monte Carlo
 & {\textemdash} & {\textemdash} & {\textemdash} & 0.02 & \\
Wulferding \textit{et al.}\ (2026)\footnotemark[7]\cite{Wulferding2026}
 & DFT$+U$;
 & 0.675 & -0.029 & 0.0638 & 0.095 & \\
\colrule
This work\footnotemark[8]
 & INS $+$ magneto-Raman; LSWT fit
 & 0.592(139) & -0.11946(80) & 0.1296(10) & 0.219(51) & \\
\end{tabular}
\end{ruledtabular}
\footnotetext[1]{The factor of two is fixed independently by the sources themselves.
Refs.~\onlinecite{Sawaoka1966,Okada1977} state the molecular-field relation
$k_{B}T_{N}=\tfrac{2}{3}S(S+1)(-4J_{1}^{\mathrm{pub}}+6J_{2}^{\mathrm{pub}})$, which under the
mapping above is $k_{B}T_{N}=\tfrac{1}{3}S(S+1)(4J_{1}-6J_{2})$ in the convention of Eq.~(2).
Evaluated with the tabulated parameters it returns $87$, $211$, $81$ and $102$~K, reproducing the
N\'eel temperatures quoted in Refs.~\onlinecite{Sawaoka1966,Okada1977,vanKalkeren1978} (the last
for both of its model parameter sets).}
\footnotetext[2]{Derived from the Curie--Weiss constant $\theta=-472$~K together with
$T_{N}=87.2$~K; the same work reports $\mu_{\mathrm{eff}}=5.84\,\mu_{B}$ and
$J_{2}/J_{1}=0.19$.}
\footnotetext[3]{$J_{1}$ was obtained from the susceptibility at the ordering temperature
via $\chi(T_{N})=Ng^{2}\mu_{B}^{2}/32|J_{1}|$, with $J_{2}$ then fitted to the
high-temperature data; the same fit returns $g=2.04(4)$. The quoted parameters give
$T_{N}=211$~K in the molecular-field approximation, against $80$~K from the
Green's-function treatment cited therein.}
\footnotetext[4]{Calculated for an effective-electron cation polarisability
$\lambda_{\mathrm{Mn}^{2+}}=0.95\,\mathrm{au}^{-1}$; the alternative $0.90\,\mathrm{au}^{-1}$
parameter set of the same work gives $J_{1}=0.655$ and $J_{2}=-0.069$~meV. That work also lists
experimental reference values of $-7.2$ and $-1.2$~K.}
\footnotetext[5]{Obtained from dilute Mn clusters in the isostructural diamagnetic host
Mg$_{1-p}$Mn$_{p}$Te$_{2}$ with $p<0.02$, not from stoichiometric MnTe$_{2}$; the same work
reports $J_{3}=0.1723$~meV. Its convention is assumed to follow Ref.~\onlinecite{Okada1977}.}
\footnotetext[6]{Illustrative model parameters, quoted in units of $J_{1}$ and chosen so as to
stabilise the observed anticollinear structure rather than fitted to data; the same work adopts
$J_{2}=-0.05\,J_{1}$ for MnTe$_{2}$ and $D_{1}=0.07\,J_{1}$ for MnS$_{2}$.}
\footnotetext[7]{Magnetic force linear-response theory applied to a DFT$+U$ electronic
structure with $U_{\mathrm{eff}}=4.2$~eV on Mn~$3d$; the reported $|\bm{D}_{1}|/J_{1}$
varies by less than $20\%$ over $U_{\mathrm{eff}}=1.0$--$5.0$~eV. That calculation also
returns symmetric anisotropic exchange $\Gamma_{1}$ components of order $10^{-3}$~meV
and $\bm{D}_{2}=0$ as symmetry required.}
\footnotetext[8]{The third-, fourth- and fifth-neighbour shells each split into two
symmetry-inequivalent bonds in $Pa\bar{3}$: $J_{3}^{(1)}=-0.05712(169)$,
$J_{3}^{(2)}=-0.08659(245)$, $J_{4}^{(1)}=-0.020076(577)$, $J_{4}^{(2)}=-0.07905(221)$,
$J_{5}^{(1)}=0.12331(364)$, $J_{5}^{(2)}=0.010620(314)$~meV.}
\end{table}

\begin{figure}[h]
 \centering
 \includegraphics[scale=0.6]{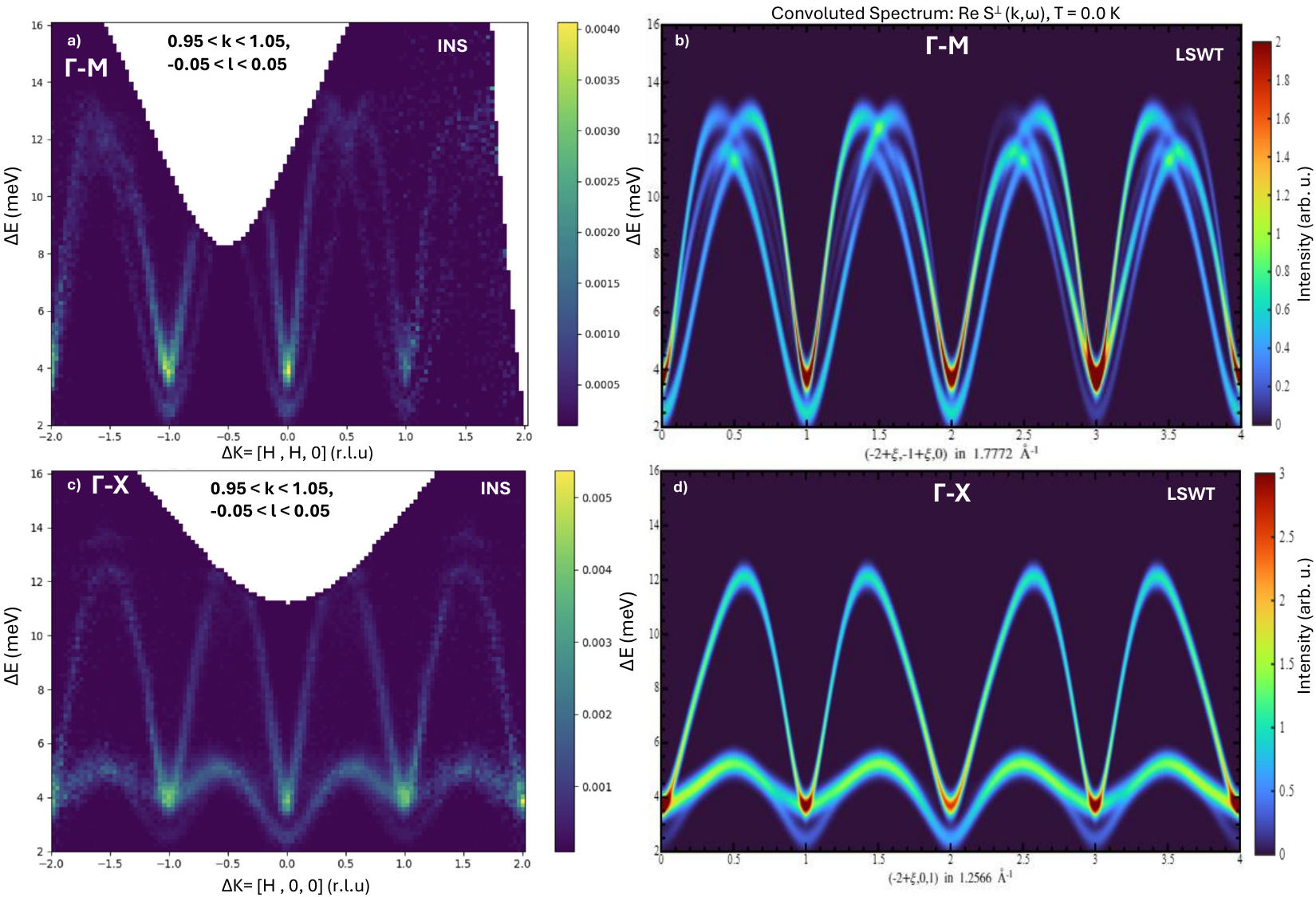}
 \caption{\textbf{Experimental and simulated INS intensity on the $\Gamma -M$ and $\Gamma -X$ cuts.} a) Experimental INS intensity on the $\Gamma-M$ line with $k,l$ taking the highlighted values. The symmetry group $Pa\bar{3}$ is assumed for the data collection. b) Simulated INS intensity on the $\Gamma-M$ line using the spin model described in the main text. c) Experimental INS intensity on the $\Gamma-X$ line with $k,l$ taking the highlighted values. d) Simulated INS intensity on the $\Gamma-X$ line. LSWT calculations are performed using the spinW program \cite{spinwToth_2015}. }
 \label{comparisonmap1}
 \end{figure}

\begin{figure}[h]
 \hspace{-1cm}
 \includegraphics[scale=0.6]{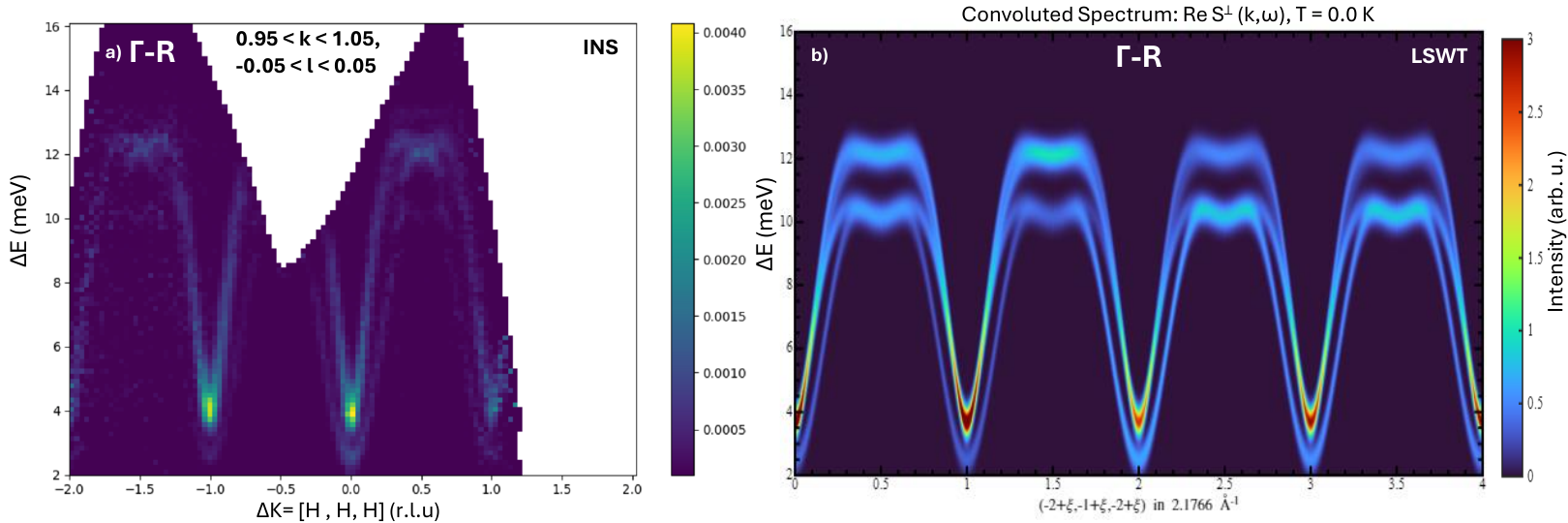}
 \caption{\textbf{Experimental and simulated INS intensity on the $\Gamma -R$ cut.} a) Experimental INS intensity on the $\Gamma-R$ line with $k,l$ taking the highlighted values. The symmetry group $Pa\bar{3}$ is assumed for the data collection. b) Simulated INS intensity on the $\Gamma-R$ line using the spin model described in the main text. }
 \label{comparisonmap2}
 \end{figure}

 \begin{figure}[h]
 \centering
 \includegraphics[width=1.\columnwidth]{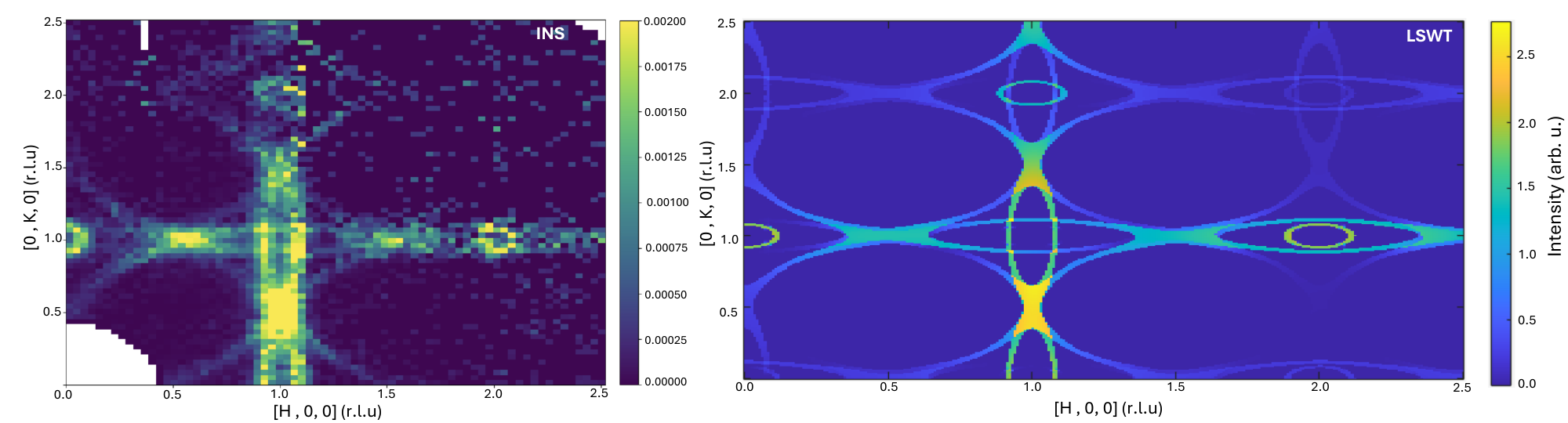}
 \caption{Experimental (left) and simulated (right) INS intensity on the $k_z=0$ plane at energy $E=5.1$ meV. INS data are collected over a rangle of $[0,0,L] \in [0,0,0.05]$ and $E \in [5.10,5.20]$, and the spinW simulated map adopts an energy window of $\pm 0.15$ meV. The symmetry group $Pa\bar{3}$ is assumed for the data collection.}
 \label{comparisonmap3}
 \end{figure}

 \begin{figure}[h]
 \hspace{-0cm}
\includegraphics[width=1.\columnwidth]{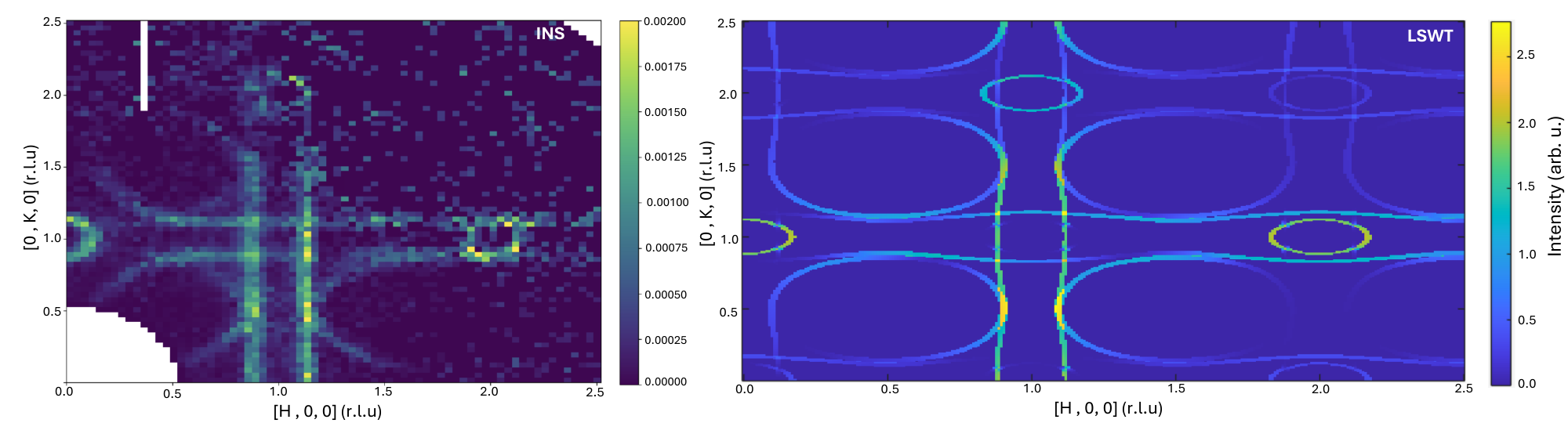}
 \caption{Experimental (left) and simulated (right) INS intensity on the $k_z=0$ plane at energy $E=6.3$ meV. INS data are collected over a rangle of $[0,0,L] \in [0,0,0.05]$ and $E \in [6.20,6.40]$, and the spinW simulated map adopts an energy window of $\pm 0.15$ meV. The symmetry group $Pa\bar{3}$ is assumed for the data collection. }
 \label{comparisonmap4}
 \end{figure}

\section{Correlation matrix of the spin-model parameters}
\label{sec:sm-corr}

The spin model carries eleven continuous parameters --- the eight Heisenberg
couplings $J_1$, $J_2$, $J_{31}$, $J_{32}$, $J_{41}$, $J_{42}$, $J_{51}$,
$J_{52}$ and the nearest-neighbour Dzyaloshinskii--Moriya (DM) vector
$\mathbf D=(D_x,D_y,D_z)$ --- fixed by fitting the calculated magnon
dispersion to the measured one. How independently the data pin down each
parameter is quantified by standard least-squares error
analysis~\cite{SPress2007}: we sample the four magnon branches
$E_n(\mathbf k)$ on a uniform zone-filling mesh, differentiate them with
respect to each parameter to build the sensitivity (Jacobian) matrix
$J_{ai}=\partial E_a/\partial\theta_i$, and obtain the parameter covariance
as the inverse of the Fisher matrix, $C\propto(J^{\top}J)^{-1}$. Its
normalization,
\begin{equation}
\rho_{ij}=\frac{C_{ij}}{\sqrt{C_{ii}\,C_{jj}}}\in[-1,1],
\label{eq:sm-corr}
\end{equation}
is independent of the overall error scale of the fitted energies.
$\rho_{ij}\to+1$ means the two parameters can drift upward together at
almost no cost to the fit (only their difference is constrained),
$\rho_{ij}\to-1$ that they trade off (only their sum is constrained), and
$\rho_{ij}\approx0$ that the dispersion fixes them independently.

\begin{figure}[tb]
\centering
\includegraphics[width=0.62\columnwidth]{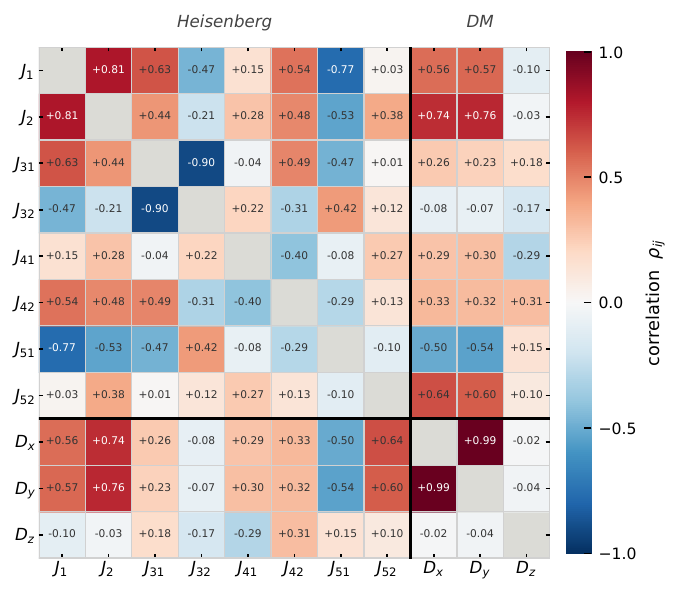}
\caption{\label{correlationmatrix}%
Correlation matrix $\rho_{ij}$ [Eq.~(\ref{eq:sm-corr})] of the eleven
spin-model parameters, from the least-squares sensitivity analysis of the
four-band $B=0$ magnon dispersion sampled on a $16^3$ 
mesh; the error scale of the fit cancels in the normalization, the trivial
unit diagonal is masked (grey), and the black lines separate the
Heisenberg and DM blocks. The large entries mark the combinations the
dispersion cannot separate --- e.g. $\rho(D_x,D_y)=+0.99$ and $\rho(J_{31},J_{32})=-0.90$ --- whereas $D_z$ and $J_{41}$ are
nearly uncorrelated with all other parameters and individually robust.}
\end{figure}

Supplementary Fig. ~\ref{correlationmatrix} shows the result of the calculated correlation matrix. (i)~The in-plane DM components
are locked, $\rho(D_x,D_y)=+0.99$: individually they carry large,
tightly-coupled uncertainties. (ii)~The two orbits of
the third-neighbor shell trade off, $\rho(J_{31},J_{32})=-0.90$: the
magnon energies fix the total third-neighbor coupling but not its split between the
orbits. (iii) The remaining sizeable entries tie the nearest-neighbor coupling to the longer-range ones,  $\rho(J_1,J_2)=+0.81$ and $\rho(J_1,J_{51})=-0.77$, implying that the magnon energies depend on the couplings only through combinations of their shells, so the dispersion constrains the overall exchange profile more tightly than the individual shell values. By contrast, e.g. $D_z$ and $J_{41}$ correlate only weakly with everything else and are individually robust. The practical
conclusion is that one-parameter error bars are misleading for
$\{D_x,D_y\}$ and $\{J_{31},J_{32}\}$: those parameters are meaningful in
the combinations the dispersion constrains, and separating them would
require additional observables.

\begin{figure}[tb]
\centering
\includegraphics[width=0.5\columnwidth]{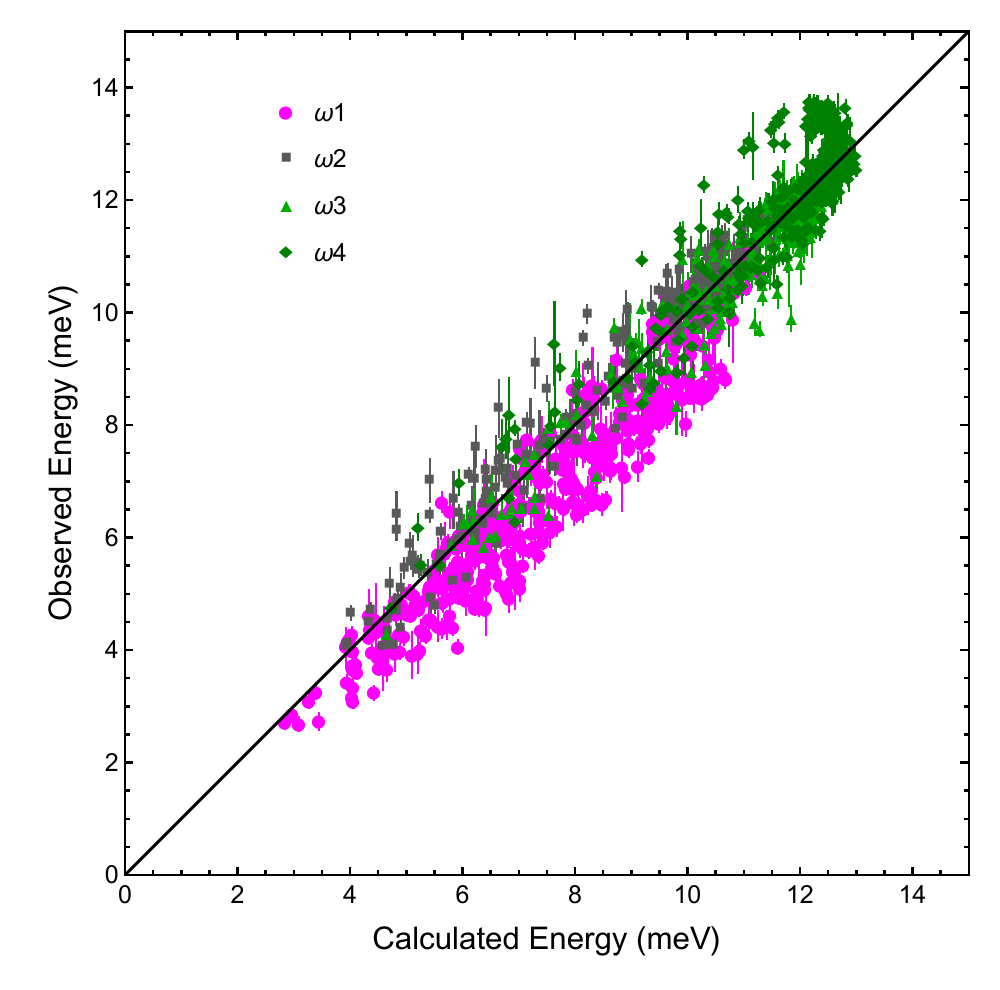}
\caption{\label{comparison_between_observed_vs_calculated} \textbf{Comparison of observed and calculated magnon energies.}
    Measured spin-wave energies extracted from the inelastic neutron
    scattering data are plotted against the corresponding energies calculated
    from the fitted spin-wave model, for magnon bands labeled
    $\omega_1$--$\omega_4$ in order of increasing energy (magenta circles,
    gray squares, dark-green up-triangles, and darker-green diamonds,
    respectively). Each observed energy was obtained by fitting a Gaussian
    lineshape to a constant-$\mathbf{Q}$ energy cut at a randomly sampled
    wavevector, and is compared to the model band whose calculated energy lies
    closest to it, with a total number of 1630 randomly chosen wavevectors. Vertical error bars represent one standard deviation in the
    fitted peak centroid, taken from the square root of the corresponding
    diagonal element of the covariance matrix of the least-squares fit, in
    which each energy bin was weighted by its intensity uncertainty. The solid black line
    denotes exact one-to-one agreement. The close adherence of the data to
    this line across the full $\omega_1$--$\omega_4$ bandwidth (within a $\pm2$ meV window) demonstrates
    that the model reproduces a good agreement to the measured dispersion throughout the Brillouin
    zone, given the experimental resolution and the possible phonon contamination discussed earlier. The left uppermost concentration of points above the black line is due to the phonon modes observed in our INS measurements at $\sim 13$ meV.  }
\end{figure}

\section{Magnon Bands Connectivity}
\label{magnon-band-connectivity-supp-section}
The enforcement of the magnon nodal loops can be explained from the connectivity of the magnon \textit{irreps} as following. Starting from the point $X (0,\pi,0)$ we have $X_1 (2)$ and $X_2(2)$ \textit{irreps} which decompose as following when moving from $X (0,\pi,0)$ to $DT (0,v,0)$:

$$
\begin{aligned}
& X_1(2) \rightarrow DT_2(1) \oplus DT_3(1) \\
& X_2(2) \rightarrow DT_1(1) \oplus DT_4(1)
\end{aligned}
$$
and as going from $DT (0,v,0)$ to $A(u,v,0)$:
$$
\begin{aligned}
& DT_1(1) \rightarrow A_1(1)  \\
& DT_2(1) \rightarrow A_2(1)  \\
& DT_3(1) \rightarrow A_2(1)  \\
& DT_4(1) \rightarrow A_1(1)  
\end{aligned}
$$
and the $A(u,v,0)$ magnon \textit{irreps} combine at $ZA(\pi,u,0)$ as:
$$
\begin{aligned}
& A_1(1) \oplus A_2(1) \rightarrow ZA_1(2) \\
\end{aligned}
$$
which, with the correct order of \textit{irreps} at $DT(0,v,0)$ and $A(u,v,0)$, results in an unavoidable crossing going from $A(u,v,0)$ to $ZA(\pi,v,0)$, giving rise to a nodal curve. Taking into account the symmetries of $Pa\bar{3}$, symmetry-related nodal lines exist on the $k_z=0$ plane and in the other $k_i=0$ planes. The nodal lines on the different $k_i=0$ planes and their momentum-energy structure are shown in Fig. \ref{mnte2_Extract_Nodal_Line} in the main text.

\section{Topology of Middle Magnon Band Gap}
\label{topo-of-middle-band-gap}
 Firstly note that the standard axes and choice of origin for both groups are the same. Then, at $\Gamma(0,0,0)$, the decomposition of group representations, upon application of external perturbations and breaking of magnetic space groups, goes as following:

$$
\begin{aligned}
\Gamma_3^{+}(1) & \rightarrow \Gamma_1^{+}(1) \\
\Gamma_4^{+}(3) & \rightarrow 3 \Gamma_1^{+}(1) 
\end{aligned}
$$

At $R\left(\frac{1}{2}, \frac{1}{2}, \frac{1}{2}\right)$, equivalent to $R\left(\frac{1}{2}, \frac{1}{2}, \frac{1}{2}\right)$ in the original $\mathcal{MSG}$,

$$
\begin{aligned}
& R_2^{+}(2) \rightarrow 2 R_1^{+}(1) \\
& R_3^{+}(2) \rightarrow 2 R_1^{+}(1) \\
\end{aligned}
$$

At $T\left(0, \frac{1}{2}, \frac{1}{2}\right)$, equivalent to $\left\{3_{111}^{+} \mid 0\right\} \boldsymbol{k}_{M\left(\frac{1}{2}, \frac{1}{2}, 0\right)}$ in the original $\mathcal{MSG}$,

$$
\begin{aligned}
& M_1(2) \rightarrow T_1^{+}(1) \oplus T_1^{-}(1) \\
& M_2(2) \rightarrow T_1^{+}(1) \oplus T_1^{-}(1)
\end{aligned}
$$

At $U\left(\frac{1}{2}, 0, \frac{1}{2}\right)$, equivalent to $\left\{3_{111}^{-} \mid 0\right\} \boldsymbol{k}_{M\left(\frac{1}{2}, \frac{1}{2}, 0\right)}$ in the original $\mathcal{MSG}$,

$$
\begin{aligned}
& M_1(2) \rightarrow U_1^{+}(1) \oplus U_1^{-}(1) \\
& M_2(2) \rightarrow U_1^{+}(1) \oplus U_1^{-}(1)
\end{aligned}
$$

At $V\left(\frac{1}{2}, \frac{1}{2}, 0\right)$, equivalent to $M\left(\frac{1}{2}, \frac{1}{2}, 0\right)$ in the original $\mathcal{MSG}$,

$$
\begin{aligned}
& M_1(2) \rightarrow V_1^{+}(1) \oplus V_1^{-}(1) \\
& M_2(2) \rightarrow V_1^{+}(1) \oplus V_1^{-}(1)
\end{aligned}
$$

At $X\left(\frac{1}{2}, 0,0\right)$, equivalent to $\left\{3_{111}^{-} \mid 0\right\} \boldsymbol{k}_{X\left(0, \frac{1}{2}, 0\right)}$ in the original $\mathcal{MSG}$,

$$
\begin{aligned}
& X_1(2) \rightarrow X_1^{+}(1) \oplus X_1^{-}(1) \\
& X_2(2) \rightarrow X_1^{+}(1) \oplus X_1^{-}(1)
\end{aligned}
$$
At $Y\left(0, \frac{1}{2}, 0\right)$, equivalent to $X\left(0, \frac{1}{2}, 0\right)$ in the original $\mathcal{MSG}$,

$$
\begin{aligned}
& X_1(2) \rightarrow Y_1^{+}(1) \oplus Y_1^{-}(1) \\
& X_2(2) \rightarrow Y_1^{+}(1) \oplus Y_1^{-}(1)
\end{aligned}
$$

Finally, at $Z\left(0,0, \frac{1}{2}\right)$, equivalent to $\left\{3_{111}^{+} \mid 0\right\} \boldsymbol{k}_{X\left(0, \frac{1}{2}, 0\right)}$ in the original $\mathcal{MSG}$,

$$
\begin{aligned}
& X_1(2) \rightarrow Z_1^{+}(1) \oplus Z_1^{-}(1) \\
& X_2(2) \rightarrow Z_1^{+}(1) \oplus Z_1^{-}(1)
\end{aligned}
$$
Out of the $10$ subgroups of $Pa\Bar{3}$, only $5$ have nontrivial $\mathcal{SI}$ groups. Among those, only one subgroup $P\Bar{1}$ $(2.4)$ host type-I topological magnons. This subgroup can be realized by e.g. a magnetic field along a low-symmetry direction. This achieves the required symmetry-breaking into $P\Bar{1}$ and the little-group \textit{irreps} at the \textit{maximal} \textit{k}-points of $P\Bar{3}$ decompose into \textit{irreps} of $P\Bar{1}$ as detailed in Methods. 
Therefore three band gaps open between the magnon bands at all high-symmetry momenta. We can then calculate the $\mathcal{SI}$ $\mathbb{Z}_2^3 \times \mathbb{Z}_4$ of each band gap using the $\mathcal{SI}$ formula of $P\Bar{1}$:
\begin{equation}
    \begin{aligned}
& z_1=n^{\Gamma_1^{-}}+n^{T_1^{-}}+n^{Y_1^{-}}+n^{Z_1^{-}} \bmod 2 \\
& z_2=n^{\Gamma_1^{-}}+n^{U_1^{-}}+n^{X_1^{-}}+n^{Z_1^{-}} \bmod 2 \\
& z_3=n^{\Gamma_1^{-}}+n^{V_1^{-}}+n^{X_1^{-}}+n^{Y_1^{-}} \bmod 2 \\
& z_4=n^{\Gamma_1^{-}}+n^{R_1^{-}}+n^{T_1^{-}}+n^{U_1^{-}}+n^{V_1^{-}}+n^{X_1^{-}}+n^{Y_1^{-}}+n^{Z_1^{-}} \bmod 4
\end{aligned}
\label{siformulas}
\end{equation}
The different orders of \textit{irreps} at the high-symmetry points result in a number of different values for each gap SI value. The possible values are summarized in Supplementary Table \ref{symmetry-indicators-table}.

{
\renewcommand{\arraystretch}{1.7} 
\begin{center}
\begin{longtable}{cc}
\caption{Possible values for the $\mathcal{SI}$ in each band gap \cite{karaki2024highthroughputsearchtopologicalmagnon} after symmetry breaking from supergroup $Pa\bar{3}$ into subgroup $P\bar{1}$. Gap $i$ refers to the band gap between the $i$-th and $(i+1)$-th lowest energy bands. Green $\mathcal{SI}s$ mark the realized values in a field of $14$ T along the low-symmetry direction $[0\; 1\; 6]$ using the spin model obtained from fitting INS and Raman data. Findings of this paper does not depend on the details of the used low-symmetry magnetic field. The field strength of $14$ T was adopted to induce a measurable middle band gap opening in future in-field INS experiments. The fourth gap refers to the collection of all four magnon bands which is always trivial. More technically, it is an elementary band representation \cite{karaki2024highthroughputsearchtopologicalmagnon,doi:10.1126/sciadv.ade7731} induced from the local ``orbitals'' $S_i^{\pm}$ 
.\label{symmetry-indicators-table}}\\
\toprule
Gap\#& Possible $\mathcal{SI}$ Values\\
\midrule
\endfirsthead
\multicolumn{2}{@{}l}{\ldots continued}\\
\toprule
Gap\#& Possible SI Values\\
\midrule
\endhead
\multicolumn{2}{r}{continued on next page\ldots}
\endfoot
\bottomrule
\endlastfoot
\multirow{6}{*}{1}&\multirow{1}{*}{0000, 0003, 0010, 0011, 0012}\\*
&\multirow{1}{*}{0013, 0100, 0101, 0102, 0103}\\*
&\multirow{1}{*}{0110, 0111, 0112, 0113, 1000}\\*
&\multirow{1}{*}{1001, 1002, 1003, 1010, 1011}\\*
&\multirow{1}{*}{1012, 1013, 1100, 1101, \textcolor{green}{1102}}\\*
&\multirow{1}{*}{1103, 1112, 1113}\\\midrule
\multirow{1}{*}{2}&\multirow{1}{*}{{\textcolor{green}{1112}}}\\\midrule
\multirow{6}{*}{3}&\multirow{1}{*}{0000, 0001, 0010, 0011, 0012}\\*
&\multirow{1}{*}{\textcolor{green}{0013}, 0100, 0101, 0102, 0103}\\*
&\multirow{1}{*}{0110, 0111, 0112, 0113, 1000}\\*
&\multirow{1}{*}{1001, 1002, 1003, 1010, 1011}\\*
&\multirow{1}{*}{1012, 1013, 1100, 1101, 1102}\\*
&\multirow{1}{*}{1103, 1111, 1112}\\\midrule
\multirow{1}{*}{4}&\multirow{1}{*}{0000}\\
\end{longtable}
\end{center}

}


\section{Field-induced symmetry lowering and the activation of the magnon
thermal Hall effect}
\label{sec:thermal-hall}

The magnetic structure of MnTe$_2$ places the four Mn$^{2+}$ moments
along the four $\langle111\rangle$ body diagonals, an arrangement invariant
under every one of the 24 operations of the paramagnetic space group $Pa\bar3$
when each is combined with the corresponding global spin rotation. The
zero-field magnetic point group is therefore the full cubic $T_h$ (type~I:
time reversal appears only through the spontaneously broken order, not as an
element of the symmetry group). The thermal Hall coefficient $\kappa_{xy}$ is the $z$ component of an
axial response vector, and within linear spin-wave theory it is carried by the
magnon Berry curvature $\Omega^{xy}_{n}(\mathbf k)$, more precisely the $z$
component of an axial vector field, through the Matsumoto--Murakami
formula~\cite{SMatsumotoMurakami2011,SMatsumotoMurakami2014}
\begin{equation}
\kappa_{xy}\;=\;-\,\frac{k_B^{2}T}{\hbar}\int_{\mathrm{BZ}}
\frac{d^{3}k}{(2\pi)^{3}}\sum_{n}
c_{2}\!\left[g(\varepsilon_{n\mathbf k})\right]\,
\Omega^{xy}_{n}(\mathbf k),
\qquad
c_{2}(\rho)=(1{+}\rho)\ln^{2}\frac{1{+}\rho}{\rho}-\ln^{2}\rho-2\,
\mathrm{Li}_{2}(-\rho),
\label{eq:sm-mm}
\end{equation}
with $g(\varepsilon)$ is the Bose function. All curvatures below are evaluated with the
paraunitary (bosonic) inner product $\langle\cdot|\Sigma_z|\cdot\rangle$
appropriate to the Bogoliubov problem~\cite{SColpa1978,STothLake2015} and
integrated by the gauge-invariant plaquette method of
Ref.~\cite{SFukui2005}.

At $B=0$ the cubic group forbids the response identically, as follows. Under a
(unitary, type-I) magnetic symmetry with rotation part $R\in O(3)$ the Berry
curvature transforms as a pseudovector field,
\begin{equation}
\vec\Omega(R\mathbf k)=\det(R)\,R\,\vec\Omega(\mathbf k),
\label{eq:sm-pseudo}
\end{equation}
so proper rotations act on $\vec\Omega$ as on a vector while inversion leaves
it even, $\vec\Omega(-\mathbf k)=\vec\Omega(\mathbf k)$. Restricting
Eq.~(\ref{eq:sm-pseudo}) to the twofold element $C_{2x}\in T_h$ and to the
transport-active component gives
$\Omega^{xy}(k_x,-k_y,-k_z)=-\Omega^{xy}(\mathbf k)$ and hence, after the
$k_z$ integration implicit in Eq.~(\ref{eq:sm-mm}),
\begin{equation}
\bar\Omega^{xy}(k_x,-k_y)=-\,\bar\Omega^{xy}(k_x,k_y),\qquad
\bar\Omega^{xy}(-k_x,k_y)=-\,\bar\Omega^{xy}(k_x,k_y)
\label{eq:sm-odd}
\end{equation}
(the second relation is from $C_{2y}$). Because $C_{2x,y}$ are exact symmetries,
the magnon energies --- and with them the $c_2$ weights of
Eq.~(\ref{eq:sm-mm}) --- coincide at partner momenta; the weighted integrand
cancels pairwise and $\kappa_{xy}\equiv0$ at every temperature.
Figure~\ref{fig:sm-maps}(a) shows the constraint~(\ref{eq:sm-odd}) realized
in the computed curvature of the lower two magnon bands: the
$k_z$-integrated curvature is odd under both reflections and
vanishes identically on the $k_x{=}0$ and $k_y{=}0$ lines, onto which the
$B=0$ nodal lines of the spectrum project.

A magnetic field enters through the Zeeman term
$H_Z=g\mu_B\mathbf B\cdot\sum_i\mathbf S_i$ with
$\mathbf B\parallel[016]$. A magnetic-group element $g=(R|\boldsymbol\tau)$
acts on the (pseudovector) spins as
$\mathbf S_i\mapsto\det(R)\,R\,\mathbf S_{g^{-1}i}$, so $H_Z$ --- and with it
the full Hamiltonian --- remains invariant if and only if
\begin{equation}
\det(R)\,R\,\mathbf B=\mathbf B ;
\label{eq:sm-survival}
\end{equation}
Applying Eq.~(\ref{eq:sm-survival}) to all 24 rotation parts of
$Pa\bar3$ singles out the field direction used in this work: $\mathbf B\parallel[016]$ --- like any direction avoiding all
symmetry axes and planes --- retains only the identity and inversion, therefore the field lowers the magnetic point group symmetry,
\begin{equation}
T_h\;\xrightarrow{\;\mathbf B\parallel[016]\;}\;C_i=\{E,i\}.
\end{equation}
Under $C_i$ the constraint structure collapses: inversion fixes
$\vec\Omega(-\mathbf k)=\vec\Omega(\mathbf k)$
[Eq.~(\ref{eq:sm-pseudo})]
but, the pseudovector $\boldsymbol\kappa$ being even under $i$, all three
components of the thermal Hall response become symmetry-allowed. The same reduction removes the protection
of the $\{100\}$-plane nodal lines, which gap everywhere except at four
isolated Weyl points.

\begin{figure}[h]
\includegraphics[width=0.95\columnwidth]{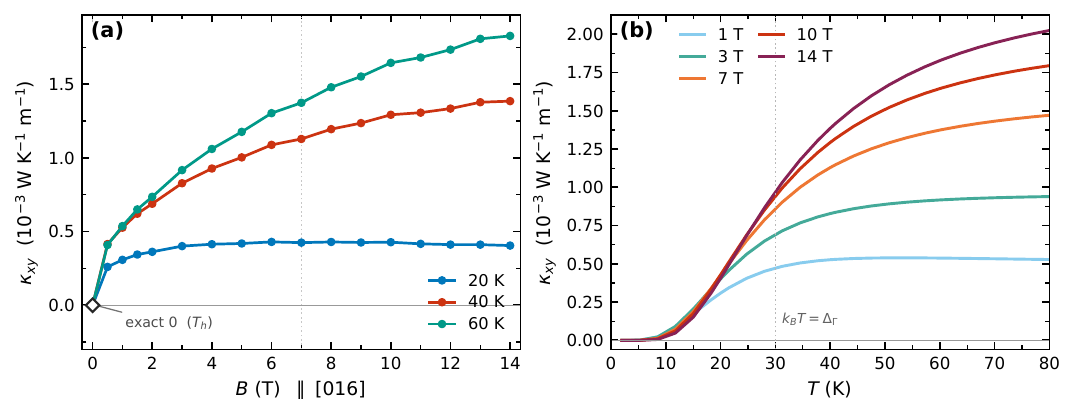}
\caption{\label{fig:sm-kappa-thermal-hall}%
Magnon thermal Hall conductivity of MnTe$_2$ for $\mathbf B\parallel[016]$
(linear spin-wave theory, $56^{3}$ Brillouin-zone grid).
(a) $\kappa_{xy}(B)$ at fixed temperatures. The $B=0$ point (open diamond) is
the exact symmetry zero enforced by $T_h$, Sec.~\ref{sec:thermal-hall}; the
response is strictly odd in $B$ and grows
monotonically through $14$~T. The dotted line marks $7$~T, the field used for
the bulk Berry-curvature figures.
(b) $\kappa_{xy}(T)$ at fixed fields: activated rise above the spin gap
($k_BT=\Delta_\Gamma\sim30$~K, dotted), followed by saturation as the $c_2$
weight approaches its high-temperature limit.}
\end{figure}

\begin{figure}[h]
\centering
\includegraphics[width=0.85\columnwidth]{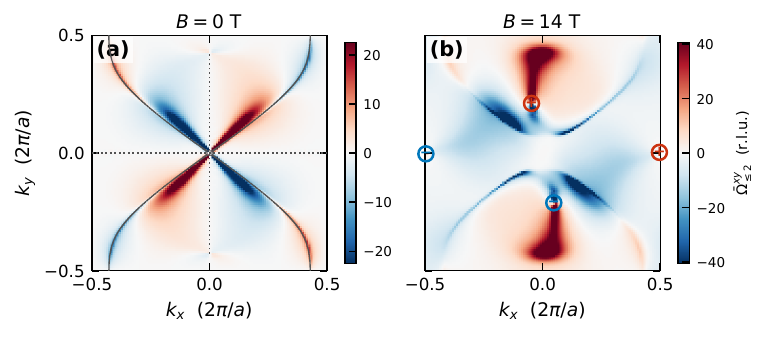}
\caption{\label{fig:sm-maps}%
$k_z$-integrated non-Abelian Berry curvature $\bar\Omega^{xy}_{\leq2}$ of the
lower two magnon bands ($112^{3}$ grid, half-step shifted so no plaquette
corner lies on a $\{100\}$ plane).
(a) $B=0$. The $2$--$3$ degeneracies form nodal lines in the $\{100\}$ planes
(solid grey: the lines lying in the $k_z{=}0$ plane; dotted grey: the
projections of the lines lying in the $k_x{=}0$ and $k_y{=}0$ planes). Each
plaquette stack pierced by a line carries its quantized $\pm\pi$ string flux
(Berry phase $\pi$). The map vanishes
identically on $k_x{=}0$ and $k_y{=}0$: the visible statement of the
cancellation that forces $\kappa_{xy}(B{=}0)=0$ at all temperatures.
(b) $B=14$~T. Only $\{E,i\}$ symmetries survive; the lines are gapped into the four Weyl
nodes (circles, colored by chirality $\chi=\pm1$), the odd symmetry is lost,
and a net integral remains.}
\end{figure}

The transport consequences are summarized in Fig.~\ref{fig:sm-kappa-thermal-hall}.
$\kappa_{xy}(B)$ switches on immediately --- the $B=0$ value is exactly zero by symmetry --- is strictly odd in $B$, and grows monotonically through $14$~T with no sign change, reaching $1.13\times10^{-3}$~W\,K$^{-1}$\,m$^{-1}$ at
$(7~\mathrm{T},40~\mathrm{K})$ and $1.39\times10^{-3}$~W\,K$^{-1}$\,m$^{-1}$
at $14$~T. The temperature dependence is as follows: the response onsets once
magnons populate the curvature-carrying low bands above the spin gap
$\Delta_\Gamma \simeq2.5$~meV ($k_B T\sim30$~K, dotted line in
Fig.~\ref{fig:sm-kappa-thermal-hall}b) and saturates as $c_2$ approaches its
high-temperature limit. 

\section{Berry-flux, curvature and the topological charges of the
field-induced Weyl nodes}
\label{berry-curvature-and-flux}

We calculate the gauge-invariant Berry
flux of the two lowest magnon bands treated as one group, evaluated with the
bosonic metric and the plaquette construction of Ref.~\cite{SFukui2005}. This
object is smooth through the internal $1$--$2$ degeneracies of the group and
singular only where band~$3$ touches band~$2$, and therefore every quantized number
below is a property of the Weyl crossings alone. Concretely, with
$\Psi(\mathbf k)$ the $2N\times2$ paraunitary eigenvector block of the two
lowest magnon branches [$\Psi^\dagger\Sigma_z\Psi=\mathbb{I}_2$,
$\Sigma_z=\mathrm{diag}(\mathbb{I}_N,-\mathbb{I}_N)$], we form on each oriented
plaquette of the discretized zone the gauge-invariant lattice field
strength~\cite{SFukui2005}
\begin{equation}
\qquad
F(\mathbf k)=\operatorname{Arg}\!\left[
U_1(\mathbf k)\,U_2(\mathbf k+\hat{\mathbf 1})\,
U_1(\mathbf k+\hat{\mathbf 2})^{-1}\,U_2(\mathbf k)^{-1}\right]
\in(-\pi,\pi], \; \text{where}\; U_\mu(\mathbf k)=\det\!\left[\Psi^\dagger(\mathbf k)\,\Sigma_z\,
\Psi(\mathbf k+\hat{\boldsymbol\mu})\right],
\label{eq:sm-fhs}
\end{equation}
whose sum over any closed discretized surface --- a constant-$k_z$ torus, or
a small sphere enclosing a node --- is $2\pi$ times an integer.

\begin{figure}[h]
\centering
\includegraphics[width=0.6\columnwidth]{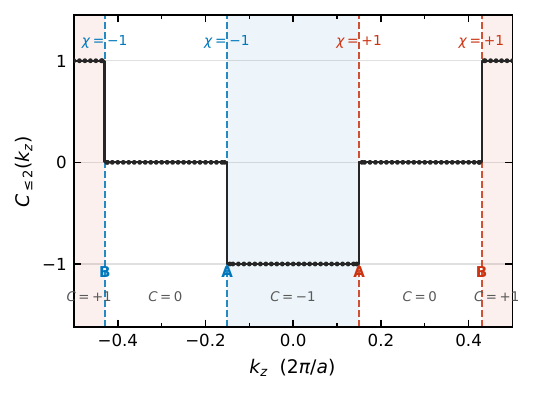}
\caption{\label{chern_number_weyls}%
The slice Chern number $C_{\leq2}(k_z)$ of
the lower two magnon bands in $k$ space at $7$~T. $C$ is integer on every gapped plane and jumps by $\pm1$ exactly at the four Weyl nodes (dashed
lines; A and B label the two inversion pairs), with each jump equal to the
sphere-flux charge $\chi$ of the node. Shading marks the three Chern phases.}
\end{figure}

Figure~\ref{chern_number_weyls} shows the Chern number $C_{\leq2}(k_z)$ of the
two-band group evaluated on constant-$k_z$ planes at $7$~T. It jumps by exactly $\pm1$ at
four values of $k_z$. Each jump coincides with a Weyl node,
and its sign equals the monopole charge $\chi$ of that node measured
independently as the Berry flux through a small enclosing sphere,
$\Phi=2\pi\chi$. The four charges form two
inversion-related pairs with $\chi=\pm1$ inside each pair and vanishing total
charge --- which is the configuration required when inversion is the only
surviving symmetry. Between the nodes every constant-$k_z$ plane is a magnon
Chern insulator threaded by quantized flux $2\pi C$; the sequence
$C=+1\rightarrow0\rightarrow-1\rightarrow0\rightarrow+1$ across the zone is
the discrete integral of the monopole field.

\begin{figure}[h]
\centering
\includegraphics[width=0.95\textwidth]{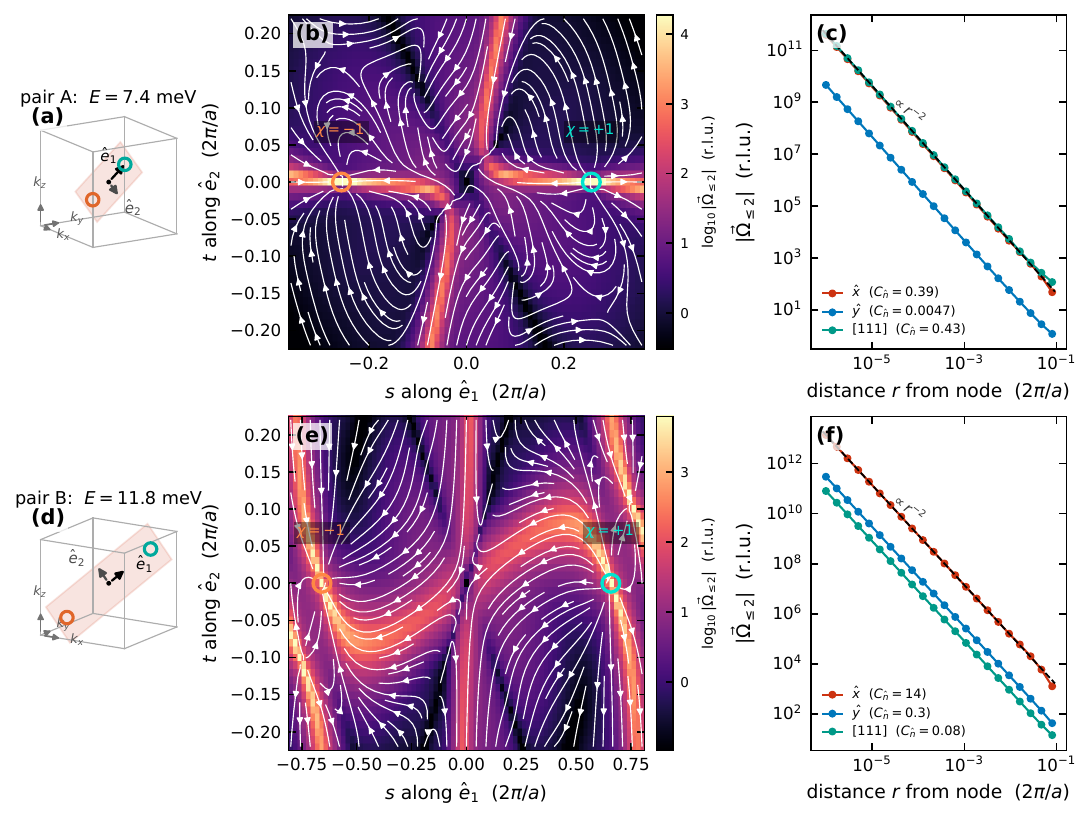}
\caption{\label{fig:sm-monopolefig}%
The Weyl nodes as Berry-flux monopoles at $7$~T (top row: pair A,
$E_A=7.38$~meV; bottom row: pair B, $E_B=11.79$~meV).
(a,d)~Location of the cut: the plane through $\Gamma$ spanned by
$\hat e_1\parallel\mathbf k_W$ and $\hat e_2$, containing both members of the
pair (circles, colored by $\chi$).
(b,e)~Streamlines of the in-plane curvature of the lower two-band group over
$\log_{10}|\vec\Omega_{\leq2}|$: a source--sink pair, with the quantitative
closed-surface statement $\oint\vec\Omega\cdot d\mathbf A=2\pi\chi$ verified
in Fig.~\ref{chern_number_weyls}.
(c,f)~$|\vec\Omega|$ along three rays from the $\chi=+1$ node: each direction
follows its own monopole law $C(\hat n)/r^{2}$
[Eq.~(\ref{eq:sm-monopole})], with measured
strengths $C_{\hat n}$ in the legends exposing the strongly anisotropic
velocity matrices of the two cones.}
\end{figure}

\emph{The monopole field.}
Near a generic (tilted, anisotropic) Weyl crossing,
$H(\mathbf q)=(\mathbf t\cdot\mathbf q)\, \mathbb{I}_2
+\sum_{ab}q_{a}V_{ab}\,\sigma_{b}$ with
$\mathbf q=\mathbf k-\mathbf k_W$, the Berry-curvature vector of the lower of
the two touching bands is exactly
\begin{equation}
\vec\Omega(\mathbf q)=-\,\frac{\det V}{2}\,
\frac{(V^{T})^{-1}\hat{\mathbf d}}{|\mathbf d|^{2}},
\qquad \mathbf d=V^{T}\mathbf q,
\qquad \chi=-\,\mathrm{sgn}\det V,
\label{eq:sm-monopole}
\end{equation}
so the tilt $\mathbf t$ drops out of the curvature entirely and the magnitude
obeys a direction-resolved monopole law
$|\vec\Omega(r\hat n)|=C(\hat n)/r^{2}$ with
$C(\hat n)=\tfrac{|\det V|}{2}\,
|(V^{T})^{-1}\widehat{V^{T}\hat n}|/|V^{T}\hat n|^{2}$.
Figure~\ref{fig:sm-monopolefig} maps this field for both pairs in the plane
through $\Gamma$ containing the $\pm\mathbf k_W$ partners: Berry flux sources
out of the $\chi=+1$ node and into its inversion partner, and along every ray
leaving a node the measured $|\vec\Omega|$ follows its own $C(\hat n)/r^{2}$
law. The measured strengths quantify a strong, pair-specific
anisotropy of the velocity matrix $V$: pair A is two orders of magnitude
softer along $\hat y$ than along its hard directions
($C_{\hat y}/C_{[111]}\simeq10^{-2}$), while pair B is concentrated along
$\hat x$ ($C_{\hat x}/C_{[111]}\simeq1.8\times10^{2}$). 

\begin{figure}[h]
\centering
\includegraphics[width=0.85\columnwidth]{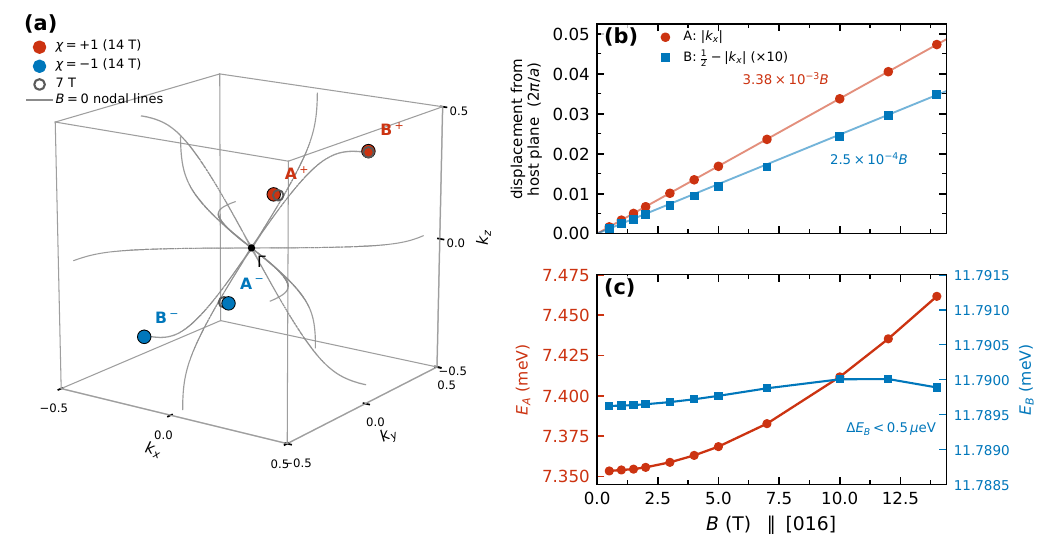}
\caption{\label{weyl-emergence}%
The field-induced Weyl nodes inside the Brillouin zone.
(a) The cubic zone with the $B=0$ nodal-line network (grey: the traced
$2$--$3$ degeneracy curves of the three $\{100\}$ planes), the four Weyl
nodes at $14$~T colored by their monopole charge $\chi=\pm1$, and their
$7$~T positions (open circles): the two inversion pairs A$^{\pm}$ and
B$^{\pm}$ are the only remnants of the network at finite field, and they
move with it.
(b) Displacement of each pair's $\chi=+1$ node from its host $\{100\}$
plane: strictly linear in $B$, with pair A leaving
the $k_x{=}0$ plane at $3.38\times10^{-3}\,(2\pi/a)$ per tesla (linear to
$0.1\%$ over the full sweep) and pair B leaving the $k_x{=}\tfrac12$ zone
face fourteen times more slowly. The $B\to0$ intercepts vanish: the nodes
are born on the parent lines, and the $B=0$ gap evaluated at the two
extrapolated birth points is $2\times10^{-4}$ and $1\times10^{-4}$~meV.
(c) Node energies: $E_A$ rises by $0.11$~meV across the sweep, while $E_B$
is pinned at $11.790$~meV to better than $0.5\,\mu$eV.}
\end{figure}

\emph{Birth and field-tuning of the nodes.}
The Weyl points do not pre-exist at $B=0$: the parent topology is a network
of $2$--$3$ nodal lines pinned to the $\{100\}$ planes, each carrying a
quantized $\pi$ Berry phase. Because the $[0 \;1\;6]$ field eliminates
symmetries that pin these lines, they gap out except where a topological
charge holds the crossings. 
%
Supplementary Fig. ~\ref{weyl-emergence} shows the resulting configuration inside the
Brillouin zone --- the four nodes as the only remnants of the parent network
--- and quantifies the field evolution of the two pairs; we track the $\chi=+1$
member of each across the field sweep. Both nodes are born on the parent lines: extrapolating
the trajectories to $B\to0$, pair A lands on the interior of the
$k_x{=}0$-plane line at $\mathbf k=(0,0.204,0.151)$ and pair B at the point
where the $k_y{=}0$-plane line pierces the $k_x{=}\tfrac12$ zone face,
$\mathbf k=(\tfrac12,0,0.428)$; the $B=0$ gap evaluated at these two birth
points is $2\times10^{-4}$ and $1\times10^{-4}$~meV. The subsequent motion is
as following: each node moves away from its host $\{100\}$ plane
strictly linearly in field while the in-plane
coordinates drift by only $\sim10^{-2}$ reciprocal-lattice units. The
monopole charges are unchanged up to $14$~T, and the node energies
behave very differently: $E_A$ rises by $0.11$~meV across the sweep whereas
$E_B$ is pinned at $11.790$~meV to better than $0.5\,\mu$eV. The field thus
acts simultaneously as the switch that creates the Weyl pairs out of
the nodal-line parent and control their positions continuously.

\begin{figure}[H]
\centering
\includegraphics[width=0.4\columnwidth]{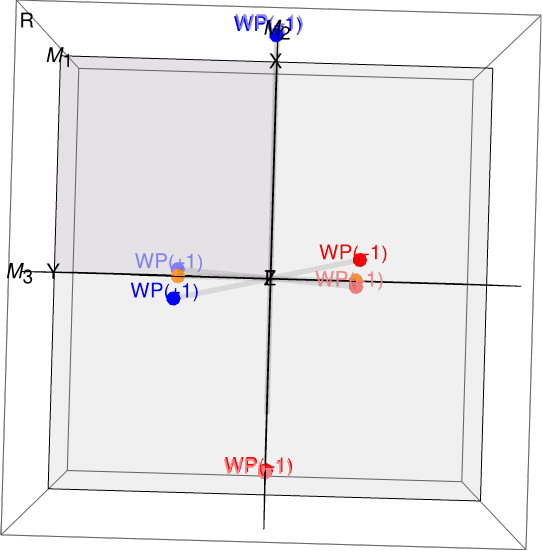}
\caption{Top view of the Brillouin zone showing the field effect on the location of the Weyl points at different magnetic field strengths applied along $[0\;1\;6]$. Bold colors correspond to $B=14~\mathrm{T}$, light colors to $B=5~\mathrm{T}$, and orange to $B=0~\mathrm{T}$ where the crossing points become part of the nodal lines, as discussed in the main text.}
\end{figure}

\section{Magnon Surface States}
\label{surface-modes}

Along $[100]$ the fcc Mn sublattice is a stack of
square-lattice atomic planes, alternating between the sublattice
compositions $\{1,4\}$ (i.e. spins at the sublattice positions $r_A$ and $r_D$) and $\{2,3\}$ (i.e. spins at the sublattice positions $r_B$ and $r_C$); cutting the crystal exposes one of
two terminations, here we adopt the termination (ending on $\{1,4\}$) throughout, and we call it the X0 termination. We compute the surface and bulk
spectral functions of the semi-infinite crystal,
\begin{equation}
A_{\rm surf/bulk}(\mathbf k_\parallel,E)=-\tfrac1\pi\,\mathrm{Im}\,
\mathrm{Tr}\,\Sigma_z\,G_{s/b},
\label{eq:sm-spec}
\end{equation}
where $G_s$ is the Green function of the outermost principal layer (four
atomic planes, grouped so that only adjacent layers couple) and $G_b$
that of a layer deep in the bulk. $G_s$ obeys the closed equation
\begin{equation}
G_s=\bigl[(E+i\eta)\,\Sigma_z-h_{00}
-h_{01}\,G_s\,h_{01}^{\dagger}\bigr]^{-1},
\label{eq:sm-sancho}
\end{equation}
with $h_{00}$ and $h_{01}$ the intra- and inter-layer blocks of the
grand dynamical matrix: the surface layer feels the rest of the crystal
only through the self-energy $h_{01}G_s h_{01}^{\dagger}$. We solve
Eq.~(\ref{eq:sm-sancho}) --- and its bulk counterpart, with self-energies
from both sides --- by the standard iterative decimation of L\'opez
Sancho \emph{et al.}~\cite{SSanchoRubio1985}, here with the bosonic
(paraunitary) metric $\Sigma_z$, and use bulk-terminated moments (the
classical surface relaxation is confined to the outermost planes and does
not move the topological features). Surface states are the
sharp, surface-dominant local maxima of $A_{\rm surf}(E)$ this criterion defines the
surface-state density of states in Supplementary Figs.~\ref{fig:sm-surfB0}(b) and
\ref{fig:sm-surf14T}(d).

\emph{$B=0$: drumhead response of the nodal-line network
(Supplementary Fig.~\ref{fig:sm-surfB0}).} The $2$--$3$ nodal network (Fig. \ref{mnte2_Extract_Nodal_Line} \& Supplementary Fig. \ref{weyl-emergence}) projects onto the (100) surface BZ: the
$k_x{=}0$-plane network maps onto itself and the other two $\{100\}$-plane
networks collapse onto axis segments [Supplementary Fig.~\ref{fig:sm-surfB0}(a)]. The
Berry phase of the lower band pair is quantized and jumps by $\pi$ across
the projected lines, so drumhead surface response is topologically
required over a large fraction of the zone. Accordingly, a surface branch closely
tracks the projected nodal-line energy [cyan overlay in Supplementary
Fig.~\ref{fig:sm-surfB0}(c)], the surface-state weight concentrates in
the upper part of the spectrum where the drumhead families live
[Supplementary Fig.~\ref{fig:sm-surfB0}(b)], and the constant-energy maps
[Supplementary Figs.~\ref{fig:sm-surfB0}(d--i)] follow the projected network as the
energy sweeps its range, from a sharp zone-center pocket to the outer
drumhead families. 

\emph{$14$~T: Weyl arcs (Supplementary Fig.~\ref{fig:sm-surf14T}).} The four field-born
Weyl nodes project onto the surface BZ at $\pm\mathbf k^\parallel_A$
($E_A$) and $\pm\mathbf k^\parallel_B$ ($E_B$). The $E_B$ map shows the
clearest arc phenomenology: sharp intensity terminating at the node
projections; on the cut through the pair-B projections the arc branch is
interrupted where the projected continuum overtakes it. The pair-A
response is resonant instead: at $E_A$ the surface enhancement near the
projections stays below the coexisting continuum, the surface-dominant
branch crosses the projection axis only above $E_A$, and the $E_A$
region contributes no surface-state weight to the DOS.

\begin{figure}[H]
\centering
\includegraphics[width=0.9\textwidth]{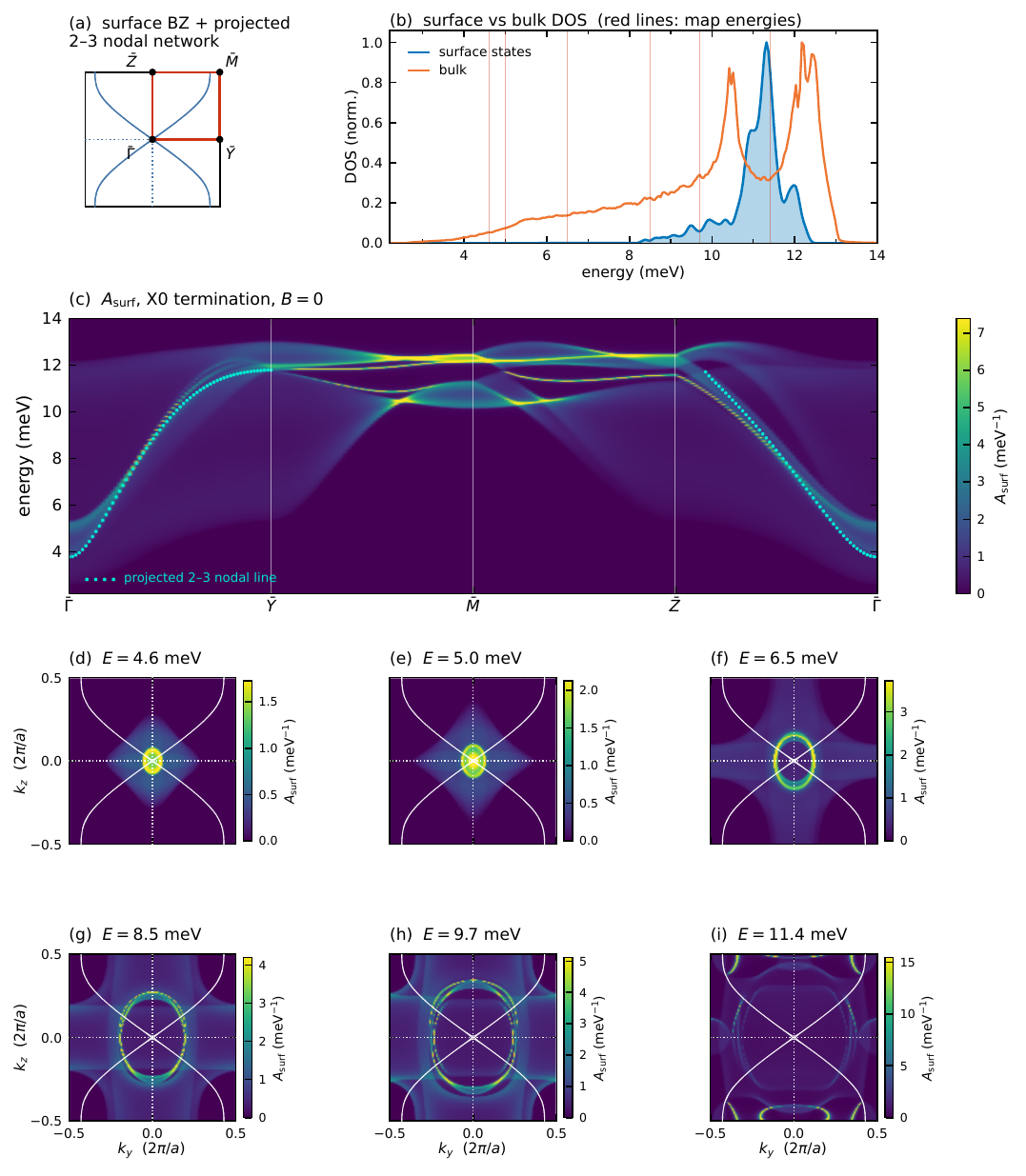}
\caption{\label{fig:sm-surfB0}%
Surface response of the nodal-line phase.
(a)~Surface BZ with the path $\bar\Gamma\bar Y\bar M\bar Z\bar\Gamma$
and the exact projection of the bulk $2$--$3$ nodal network: the
$k_x{=}0$-plane network projects onto itself (solid), while the
$k_y{=}0$- and $k_z{=}0$-plane networks collapse onto segments of the
surface-BZ axes (dotted).
(b) Surface-state DOS against the bulk DOS, each curve normalized to
its own maximum. ``Surface states'' here are the sharp, surface-dominant
local maxima of $A_{\rm surf}(E)$ lying in local gaps of the projected
bulk spectrum (specifically, we adopt the criterion $A_{\rm surf}>3A_{\rm bulk}$ with $A_{\rm bulk}<0.5$),
collected over the full surface BZ; the bulk curve is the BZ-averaged
interior DOS. The surface-state weight is confined to the drumhead
range, concentrated at $10.5$--$12.6$~meV, and vanishes at low energies
--- the low-energy surface enhancement of panel (c) is resonant and does
not qualify the criterion. Red lines mark the map energies of (d--i).
(c)~$A_{\rm surf}$ along the path ($\eta=0.015$~meV). The drumhead ribbon rises along
$\bar\Gamma\bar Y$ (mirrored on $\bar Z\bar\Gamma$), closely tracking
the projected nodal-line energy (cyan overlay); the flat, intense
families around $\bar Y\bar M$ and $\bar M\bar Z$ at
$10.8$--$12.1$~meV are surface states in the bulk DOS gap; the fainter enhancement below $\sim7$~meV is resonant.
(d--i)~Constant-energy maps ($\eta=0.010$~meV) with the
projected network of (a) overlaid. The sharp surface weight intersects the
network throughout and follows it as the energy sweeps the nodal line's
own energy range: a sharp pocket at the zone center at $4.6$--$5.0$~meV
(where the nodal line energy is lowest), an expanding ring threading the
network crossing at $6.5$--$9.7$~meV, and the outer, face-adjacent
drumhead families at $11.4$~meV.}
\end{figure}

\begin{figure}[H]
\centering
\includegraphics[width=0.9\textwidth]{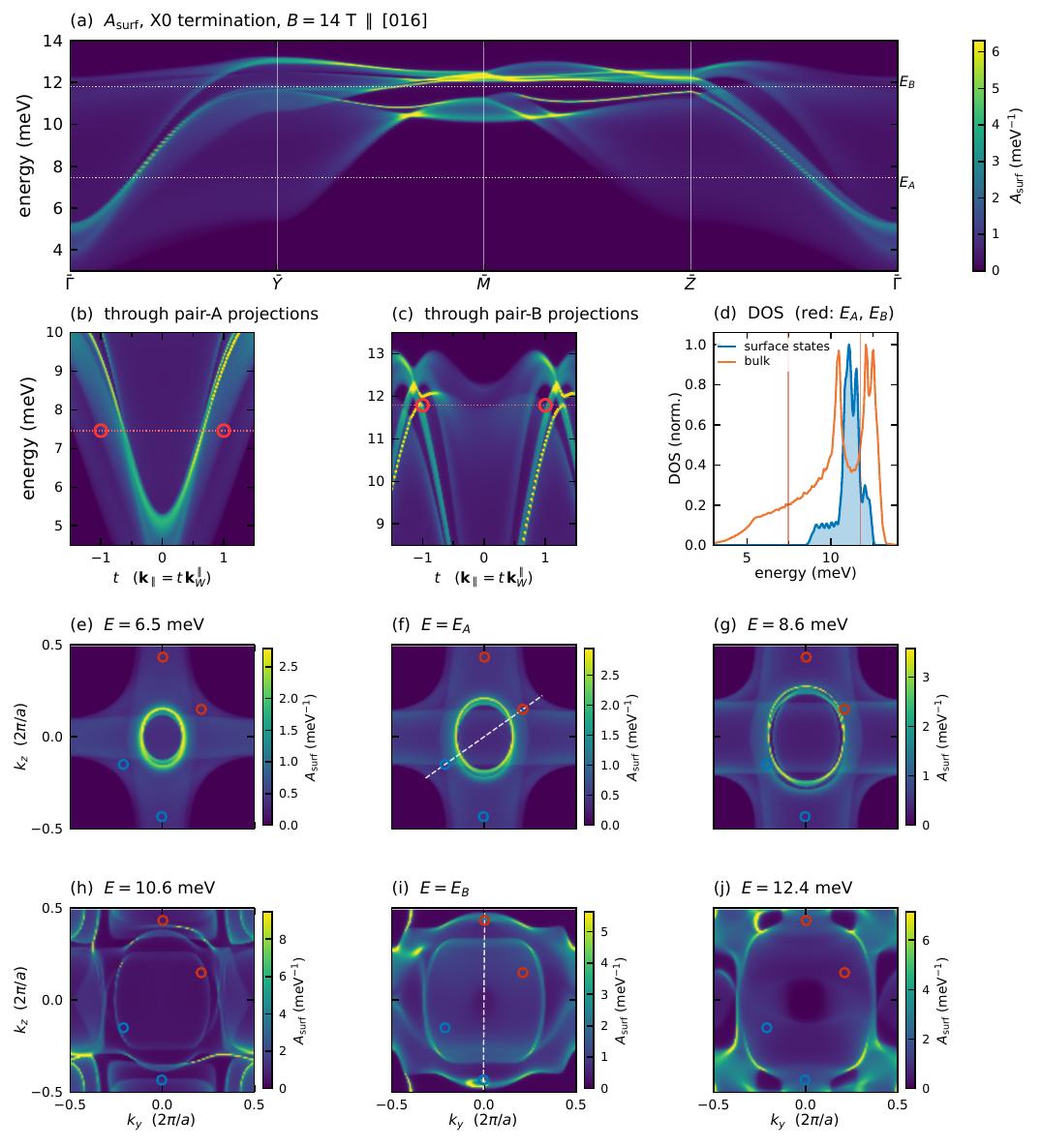}
\caption{\label{fig:sm-surf14T}%
Surface response of the Weyl phase.
(a)~$A_{\rm surf}$ along the path; dotted lines mark the Weyl energies
$E_A=7.462$ and $E_B=11.790$~meV. The arc-related families appear as the
flat intense weight at $10.8$--$12.0$~meV around $\bar Y\bar M$ and the
branch descending from $11.6$ to $8.9$~meV on $\bar Z\bar\Gamma$; near
$E_A$ the enhancement remains comparable to the continuum (resonant).
(b,c)~Spectra along the straight cuts
$\mathbf k_\parallel=t\,\mathbf k^\parallel_W$ through the pair-A and
pair-B projections [dashed lines in (f,i); circles at $t=\pm1$, $E=E_W$;
yellow: extracted surface modes]. In (b) the surface-dominant branch
crosses the projection axis at $8.3$--$10.0$~meV, \emph{above} $E_A$; at
$E_A$ itself the response stays below the coexisting continuum --- the
pair-A arc is a resonance. In (c) the arc branch passes through the
pair-B projections at $E_B$ but is interrupted on $t\in(-1,0)$, where
the projected continuum overtakes it.
(d) Surface-state DOS against the bulk DOS, with ``surface states''
defined as in Fig.~\ref{fig:sm-surfB0}(b); red lines mark $E_A$ and
$E_B$. The weight concentrates at $10.5$--$12.6$~meV --- the
field-evolved descendants of the drumhead families, now carrying the
arcs --- while the $E_A$ region contributes none, consistent with the
resonant character of pair A.
(e--j)~Constant-energy maps with all four node projections marked
(circles colored by the monopole charge $\chi$; dashed lines: the cuts
of (b,c)). The series assembles the arc weight through the Weyl window:
at $6.5$~meV the central contour has not yet reached the pair-A
projections; at $E_A$ and $8.6$~meV it passes through and beyond them
with faint endpoints; by $10.6$~meV the sharp weight has transferred to
the outer contours, which at $E_B$ (i) terminate on the pair-B
projections with a pronounced $\pm$ asymmetry
($A_{\rm surf}\simeq19$ near $-\mathbf k^\parallel_B$ against
$\simeq3$ near $+\mathbf k^\parallel_B$) --- allowed because the
$[016]$ field leaves no symmetry connecting $\pm\mathbf k_\parallel$ on
one face; at $12.4$~meV, above $E_B$, only face-adjacent remnants
survive.}
\end{figure}

 \section{DFT Phonon Calculation}
In order to help identify phonon modes and disambiguate them from magnon modes at similar energies, we perform density functional theory (DFT) \cite{DFT-Hohenberg,DFT-Kohn} calculations using the VASP code \cite{VASP-1996,VASP-1999,VASP-2006} (as a check we also find very similar results with Quantum Espresso \cite{giannozzi2009quantum}). We use a $6\times6\times6$ k-point grid, 450 eV cutoff energy, and the Mn\_sv\_GW and Te\_GW PAW files. We perform DFT calculations using the PBEsol \cite{PBEsol} functional and using the DFT+U framework \cite{DFTU}, testing U values between 1-5 eV on the Mn-3d states. U values in the range of 3-5 eV typically improve results for 3d transition metal states, which are otherwise overly delocalized due to self-interaction errors in semilocal functionals. Calculations are performed using the experimental antiferromagnetic non-collinear spin structure and with spin-orbit coupling. Phonon calculations are performed in a finite difference framework using the Phonopy code \cite{Phonopy2023,RevPhonopy2023}.

\begin{figure}[H]
    \centering \includegraphics[width=.5\linewidth]{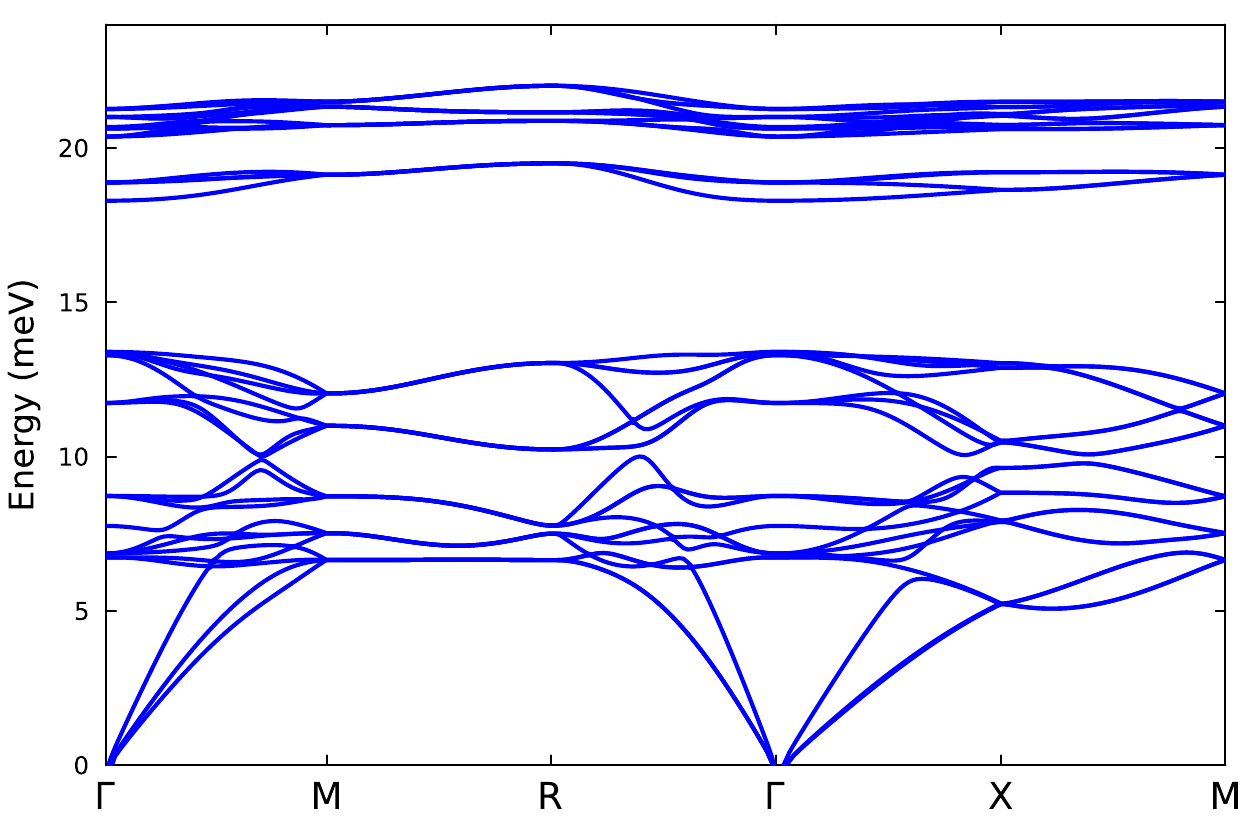}
    \caption{\textbf{DFT Calculation of Phonon Dispersion.} Full dispersion of phonons along high symmetry directions. In the range from 10 meV to 15 meV, there are phonons in the M-R momentum direction that overlap with magnons.}
    \label{DFT Phonons}
\end{figure}

We find the best agreement overall with the experimental Raman measurements are at $U =5$ eV with the experimental volume. However, $U=3-5$ eV calculations with calculated volumes are broadly in agreement with Raman data as well. We note that these calculations do not include possible coupling between phonon and magnon modes. In Supplementary Fig. \ref{DFT Phonons}, we show the phonon dispersion, in order to compare with the neutron data and LSWT fit in Figs. 1-2 of the main text. Due to the $12$ atoms in the primitive unit cell, there are a large number of phonon modes, many of which overlap with the energy range of the magnons. In particular, we note the phonon bands from $10-14$ meV which can account for the extra non-magnon bands in the Fig. 2. For example, the phonon modes from M-R at $10-13$ meV match the bands above and below the magnon bands in Fig. 2.

\section{Co-existence of phonon modes within the same magnon energy range $\lesssim 14 $ meV}
\label{phonon-high-temp-section}
\begin{figure}[H]
    \hspace{-1cm} \includegraphics[width=1.1\linewidth]{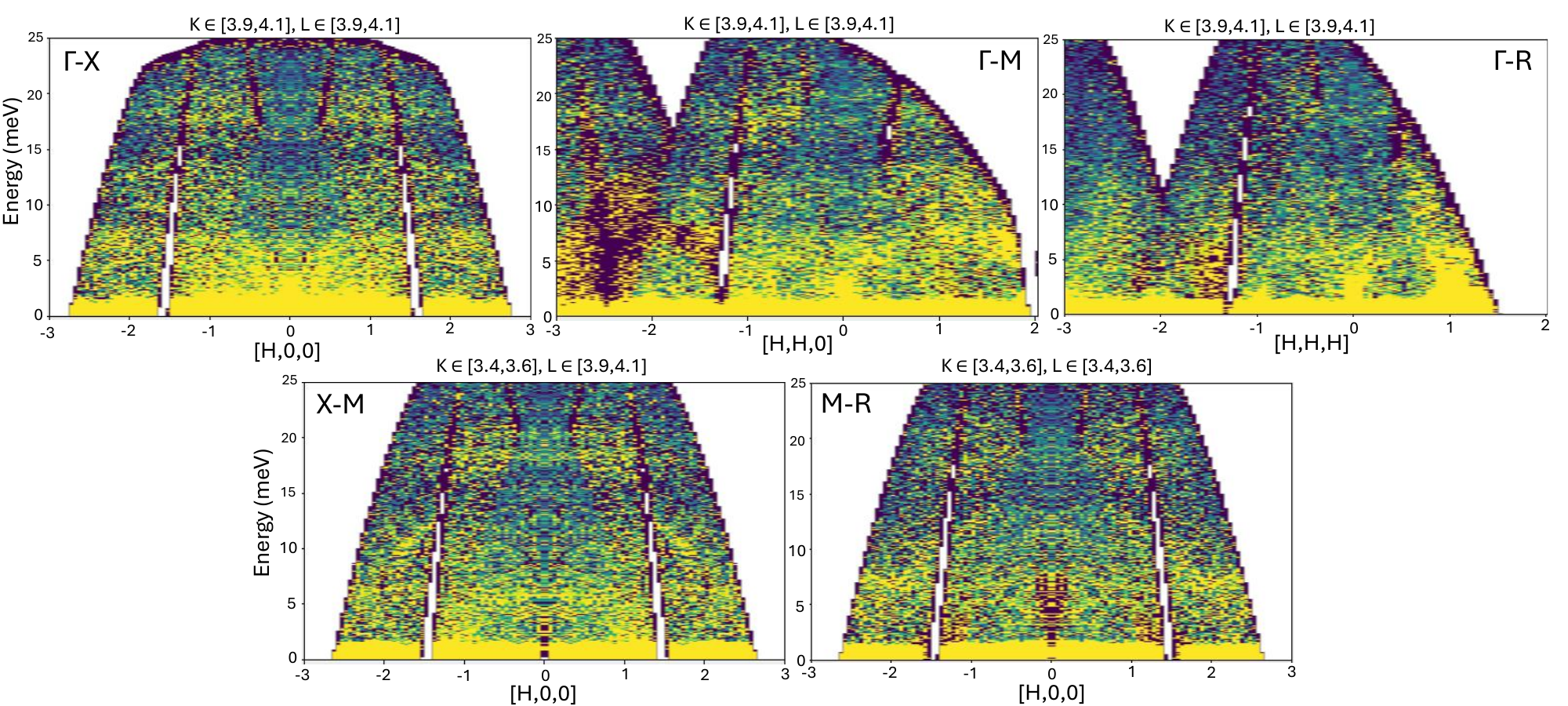}
    \caption{\textbf{High-temperature INS to probe phonons.} INS data collected at a temperature $T=105$ K, above the magnetic ordering temperature $T_N =86.5$ K, along different high-symmetry momentum cuts show the existence of phonon modes up to $\lesssim 14 $ meV and another packet of modes around $\sim [16-22]$ meV. The symmetry group $Pa\bar{3}$ is assumed for the data collection. }
    \label{phonons_INS}
\end{figure}

To reveal the phonon band structure, we performed another INS experiment at a temperature $T=105$ K well above the magnetic ordering temperature $T_N \sim 86.5$ K. Supplementary Figure \ref{phonons_INS} shows the collected INS data along the different high-symmetry momentum cuts where we observe the concentration of phonons in two energy packets. The first is up to $\sim 14$ meV while the second is $\sim [16-22]$ meV, in a rough agreement with the DFT calculated spectrum of Supplementary Figure \ref{DFT Phonons}. Even though with the weak resolution to fully resolve the shape of the phonon bands, the co-existence of the broad phonon modes with magnons within $\lesssim 14 $ meV might provide some explanation for the extra discontinuous modes $\sim 13-14$ meV observed in the low-temperature spectrum, e.g. along $X-M$ and $M-R$ in Fig. \ref{figure2fittings} and for the extra caps at $\sim 13$ meV on the $[H,0,-1-\eta]$ cuts as shown in Figure \ref{mnte2_Extract_Nodal_Line}. This might also explain the origin of the avoided-crossing points observed in our low-temperature INS data $\sim 9$ meV close to the $M$ point along $\Gamma-M$ and around the middle of $\Gamma-R$, see Fig. \ref{figure2fittings}.

 \section{Raman Polarization and Temperature Dependence}
 \label{Raman-tensors-polarization-temp}
 With the point group m$\bar{3}$, the Raman active modes are: A$_{\text{g}}$, E$_{\text{g}}$ (note that E$_{\text{g}}$ is not an irreducible representation but here we are combining two irreducible representation into a single E$_{\text{g}}$) and T$_{\text{g}}$, with the Raman tensors:\par

$$\text{A}_{\text{g}}: \begin{bmatrix}
a & 0 & 0\\
0 & a & 0\\
0 & 0 & a
\end{bmatrix}
$$

$$\text{E}_{\text{g}}: 
\begin{bmatrix}
b & 0 & 0\\
0 & b\text{e}^{-\frac{2\pi i}{3}} & 0\\
0 & 0 & b\text{e}^{\frac{2\pi i}{3}}
\end{bmatrix} \text{and} \begin{bmatrix}
b & 0 & 0\\
0 & b\text{e}^{\frac{2\pi i}{3}} & 0\\
0 & 0 & b\text{e}^{-\frac{2\pi i}{3}}
\end{bmatrix}
$$

$$\text{T}_{\text{g}}: 
\begin{bmatrix}
0 & c & 0\\
c & 0 & 0\\
0 & 0 & 0
\end{bmatrix} \text{,} \begin{bmatrix}
0 & 0 & c\\
0 & 0 & 0\\
c & 0 & 0
\end{bmatrix}
\text{and}
\begin{bmatrix}
0 & 0 & 0\\
0 & 0 & c\\
0 & c & 0
\end{bmatrix}
$$
The theoretical response at zero external magnetic field and parallel polarizations are calculated and used to fit the experimental data. When the incoming and outgoing polarization rotate in $bc$ plane, the Raman response for the phonons are: A$_{\text{g}}$:$a^2$, E$_{\text{g}}$: $2b^2(\sin ^4 \theta+\cos ^4 \theta -\sin^2\theta\cos^2\theta)$, T$_{\text{g}}$:$c^2\sin^2 2\theta$, where $a$, $b$, and $c$ come from the Raman tensors above. The Raman response for four phonons are measured and compared to the theoretical calculations, and the symmetries are identified as, from low energy to high energy, T$_{\text{g}}$, E$_{\text{g}}$, A$_{\text{g}}$, and T$_{\text{g}}$ shown in Supplementary Fig. \ref{RamanPolarizaions} as panels (a), (b), (c), and (d), respectively. Notice that the Raman response for a T$_{\text{g}}$ phonon should have nodes but there is not one in the experimental results for the highest energy phonon. This is likely due to the small scattering signal from this phonon (see Fig. 1 in the main text). The polarization response for the magnons is also shown in this figure, as panels (e) and (f), where their symmetries are identified as E$_{\text{g}}$ and T$_{\text{g}}$ in order of increasing energy.

 \begin{figure}[h]
    \centering    \includegraphics[width=0.5\linewidth]{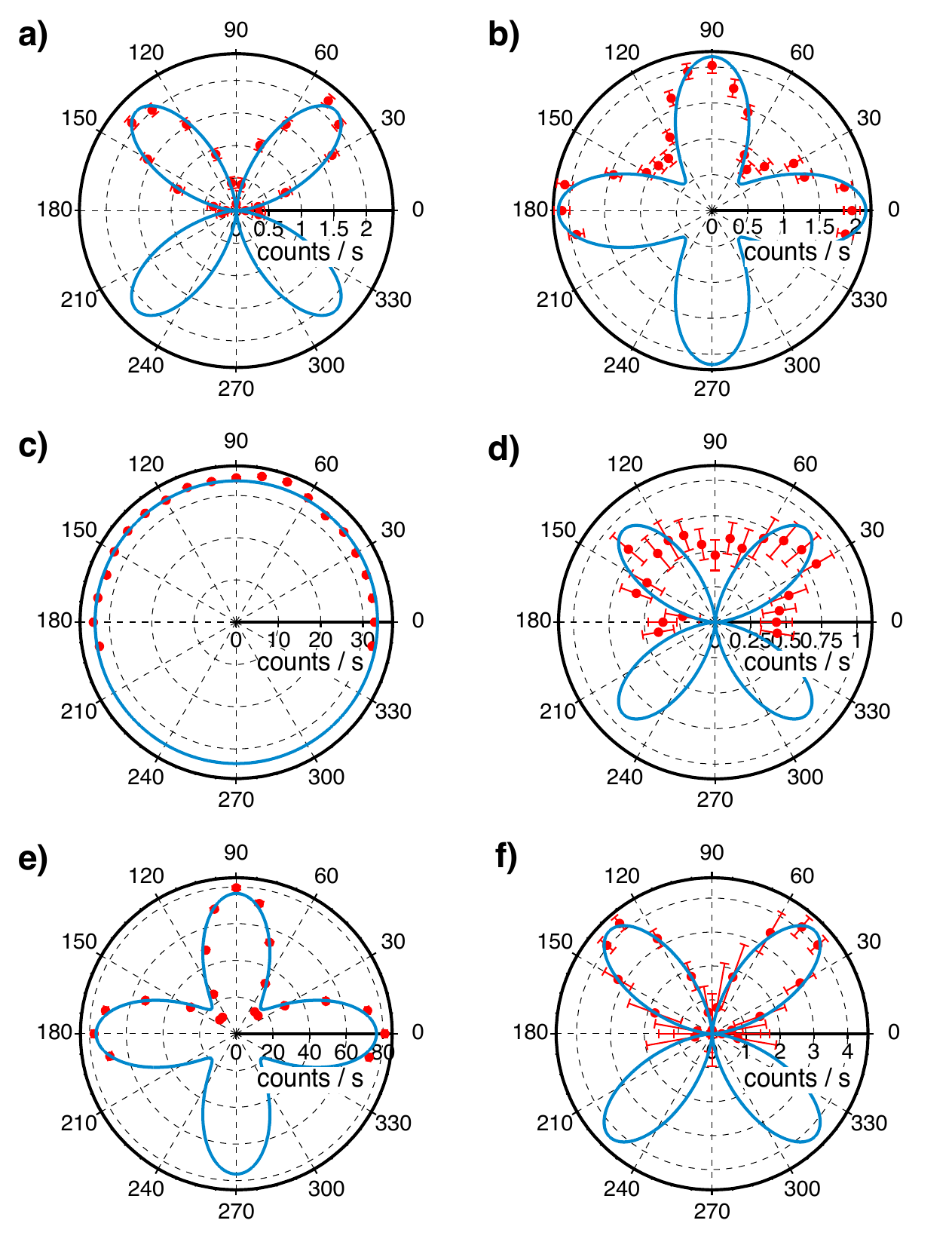}
    \caption{\textbf{Raman Polarization dependence} (dots: experimental result; lines: theoretical fit) in parallel polarizers configuration of the Raman response of the phonons (a), (b), (c), and (d) in order of increasing energies, as shown in Fig. 1, and magnons (e) and (f) in Fig. 1.}
    \label{RamanPolarizaions}
\end{figure}

The temperature dependence of the magnon energies and full width at half maxima are shown in Supplementary Fig. \ref{TempDependenceFull}. As expected, the magnon energies soften (decrease) and their widths increase as the temperature increases toward the Neel temperature.

 \begin{figure}[h]
    \centering    \includegraphics[width=.9\linewidth]{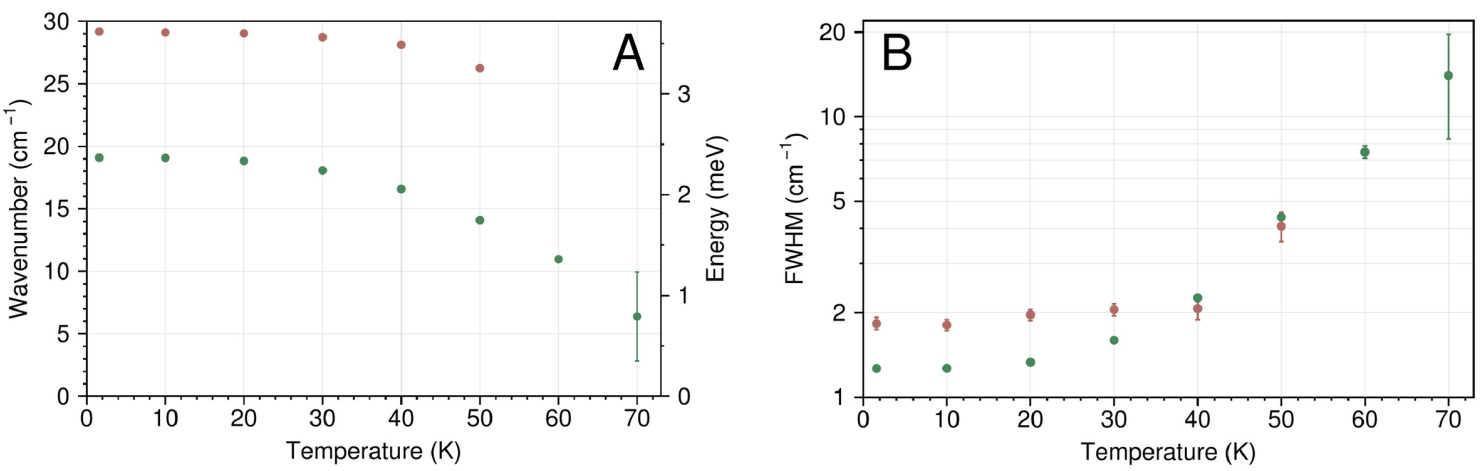}
    \caption{\textbf{Temperature dependence of $\Gamma$-magnons measured by Raman spectroscopy.} A) Energies of magnon bands as a function of temperature. The red (green) dots denote the energy of the measured triplet (singlet) magnon. B) FWHM of the triplet and singlet magnon bands as a function of temperature. As $T_N$ is approached, the energies of the magnon modes centers decrease and have broader bandwidths.}
    \label{TempDependenceFull}
\end{figure}

To figure out the direction of the magnetic field along low symmetry directions, photographs of the crystals were used to determine the cubic axes following striations of the crystal along the cubic axes (1,0,0), (0,1,0), and (0,0,1). In the Faraday configuration, where the magnetic field is applied along the direction of propagation of the outgoing photon, $k_{out}\parallel B$, a crystal was polished with a wedge under it to create a surface not perpendicular to any of the cubic axes. The direction perpendicular to this surface was determined as shown in panel a) of Supplementary Fig. \ref{samplesorientation}, using simple trigonometry by inspecting the angles that the cubic directions striations make to the wedged surface. In the Voigt configuration, where the applied magnetic field is perpendicular to the direction of propagation of the outgoing photon, $k_{out}\perp B$, a crystal was rotated around the (1,0,0) axis by a small angle and the vertical direction was chosen as the direction of the magnetic field. The angles were determined similarly to the Faraday case using simple trigonometry of the striations of the crystal. This is shown in panel (b) of Supplementary Fig. \ref{samplesorientation}.

\begin{figure}[h]
    \centering \includegraphics[width=.75\linewidth]{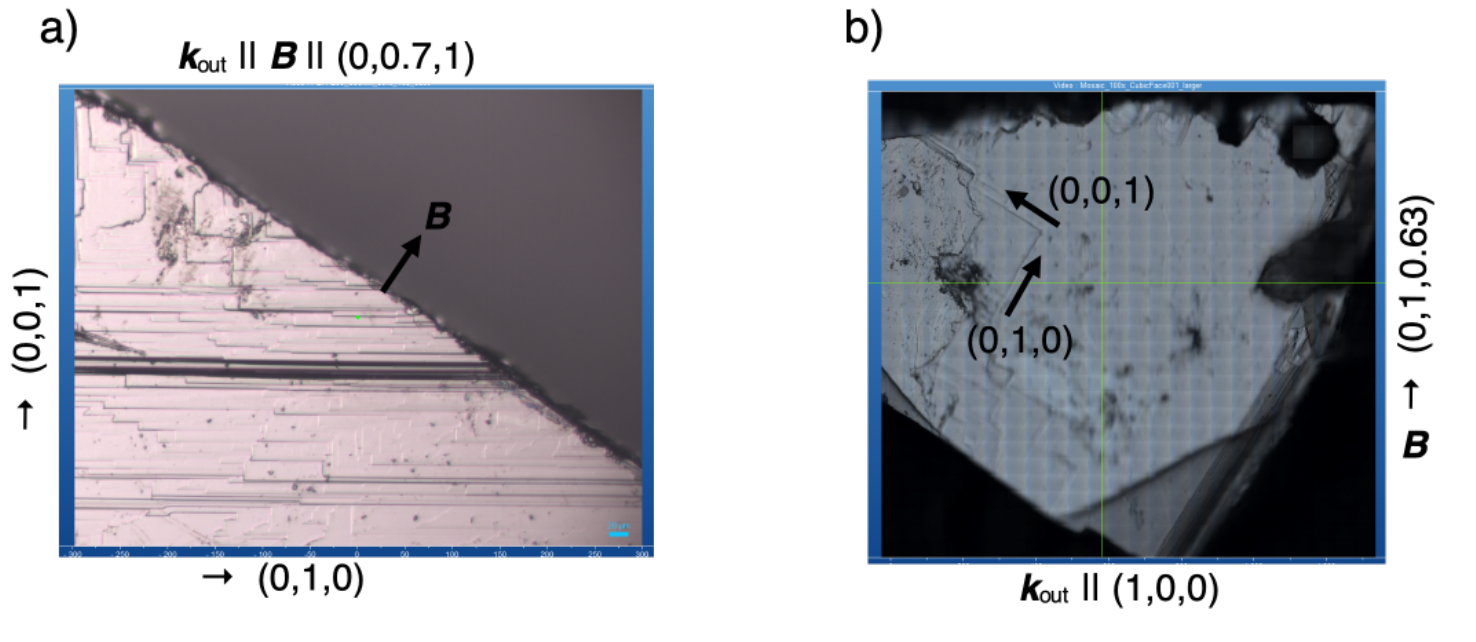}
    \caption{\textbf{Determination of magnetic field direction in the two crystals used in the Raman experiments.} Samples used in the Faraday a) and Voigt b) configurations experiments, with the cubic axes labeled and the magnetic field directions marked.}
    \label{samplesorientation}
\end{figure}

\end{document}